\documentclass[a4paper,11pt]{article}
\pdfoutput=1 
\usepackage{jheppub,slashed,bbold}

\DeclareMathOperator\Tr{Tr}
\DeclareMathOperator\diag{diag}
\newcommand{\der}{\partial}
\newcommand{\SU}{\text{SU}}
\newcommand{\SO}{\text{SO}}
\renewcommand{\O}{\text{O}}
\newcommand{\U}{\text{U}}
\newcommand{\PF}{P_F}
\newcommand{\YM}{\textrm{g}}
\newcommand{\Luv}{\Lambda_{\rm UV}}
\newcommand{\ee}{\,\mathrm{e}}
\renewcommand{\epsilon}{\varepsilon}
\renewcommand{\bar}[1]{\overline{#1}} 
\newcommand{\1}{\mathbb{1}}
\newcommand{\adj}{\text{adj}}
\newcommand{\qcdad}{{\rm QCD(adj)}}
\newcommand{\ZZ}{\mathbb{Z}}
\newcommand{\RR}{\mathbb{R}}
\renewcommand{\SS}{S}
\newcommand{\NfD}{N_f^D}
\newcommand{\NfW}{N_f^W}
\newcommand{\md}{m_{\rm dyn}}
\newcommand{\LQCD}{\Lambda_{\rm QCD}}
\newcommand{\kakko}[1]{\left\|{#1}\right\|}

\title{\boldmath 
Adjoint QCD on $\RR^3\times \SS^1$  
with twisted fermionic boundary conditions
}

\author[a]{Tatsuhiro Misumi}
\author[b]{and Takuya Kanazawa}

\affiliation[a]{Department of Physics, and Research and Education Center for Natural Sciences, 
Keio University, Hiyoshi 4-1-1, Yokohama, Kanagawa 223-8521, Japan}
\affiliation[b]{iTHES Research Group and Quantum Hadron Physics Laboratory, RIKEN, Wako, Saitama 351-0198, Japan}

\emailAdd{misumi@phys-h.keio.ac.jp}
\emailAdd{takuya.kanazawa@riken.jp}

\preprint{RIKEN-QHP-152}

\abstract{%
We investigate QCD with adjoint Dirac fermions on $\RR^3\times \SS^1$ with generic 
boundary conditions for fermions along $\SS^1$.  
By means of perturbation theory, semiclassical methods and a chiral effective model, 
we elucidate a rich phase structure in the space spanned by the $\SS^1$ compactification scale $L$, 
twisted fermionic boundary condition $\phi$ and the fermion mass $m$. 
We found various phases with or without chiral and center symmetry breaking, 
separated by first- and second-order phase transitions, which in specific limits 
($\phi=0$, $\phi=\pi$, $L\to0$ and $m\to \infty$) reproduce known results in the literature. 
In the center-symmetric phase at small $L$, we show that \"{U}nsal's bion-induced 
confinement mechanism is at work but is substantially weakened at $\phi\ne 0$ 
by a linear potential between monopoles.    
Through an analytic and numerical study of 
the PNJL model, we show that the order parameters for center and chiral 
symmetries (i.e., Polyakov loop and chiral condensate) are strongly intertwined at $\phi\ne 0$. 
Due to this correlation, a deconfined phase can intervene between a weak-coupling 
center-symmetric phase at small $L$ and a strong-coupling one at large $L$. 
Whether this happens or not depends on the ratio of the dynamical 
fermion mass to the energy scale of the Yang-Mills theory.  
Implication of this possibility for resurgence in gauge theories is briefly discussed.   
In an appendix, we study the index of the adjoint Dirac operator on $\RR^3\times \SS^1$ 
with twisted boundary conditions, which is important for semiclassical analysis of monopoles.  
}

\begin{document}
\maketitle
\flushbottom


\section{Introduction}
\label{sec:Intro}

Compared to QCD with quarks in the fundamental representation of the gauge group, 
$\SU(N)$ gauge theories with adjoint fermions [\qcdad] have many distinctive features that make {\qcdad} an important area of research. 
Firstly, adjoint fermions do not break the center symmetry $\ZZ_N$ and there is a well-defined deconfinement transition in {\qcdad}.  
In lattice simulations the deconfinement transition was found to occur at a much lower temperature than the chiral phase transition  \cite{Kogut:1985xa,Kogut:1986jt,Karsch:1998qj,Engels:2005te}. This is in contradistinction to QCD with fundamental quarks, where the two transitions (crossovers) occur at around the same temperature. Understanding the origin of such discrepancy may bring us new insights into the mechanism of confinement and chiral symmetry breaking in QCD and {\qcdad}. 

Secondly, {\qcdad}  is free from the notorious sign problem even at nonzero chemical potential and can be simulated with standard lattice QCD methods  \cite{Alford:1998sd,Kogut:2000ek,Hands:2000ei}. This peculiar feature originates from a special anti-unitary symmetry of the Dirac operator in the adjoint representation.  Owing to this symmetry, the pattern of chiral symmetry breaking is not the standard one $\SU(\NfD)_{\rm R}\times\SU(\NfD)_{\rm L}\to \SU(\NfD)_{\rm V}$, but rather $\SU(\NfW)\to\SO(\NfW)$ \cite{Leutwyler:1992yt}, where $\NfD$ is the number of Dirac fermions and $\NfW$ ($=2\NfD$) is that of Weyl fermions. Part of the Nambu-Goldstone (NG) modes are diquarks, and their Bose-Einstein condensation (BEC) at nonzero chemical potential has been studied in chiral perturbation theory \cite{Kogut:2000ek,Splittorff:2000mm,Kanazawa:2011tt} and in chiral effective models \cite{Zhang:2010kn,Braun:2012zq}. A relativistic analogue of the BEC-BCS crossover from low to high density is also conjectured \cite{He:2013gga}. These developments in {\qcdad} make it an attractive laboratory for methods and concepts of finite-density QCD. 

Thirdly, {\qcdad} with a single Majorana fermion is nothing but the $\mathcal{N}=1$ supersymmetric Yang-Mills (SYM) theory and its  
non-perturbative dynamics has been studied for long \cite{Witten:1982df}, serving as an archetype of strongly-coupled supersymmetric gauge theories. 

Other topics related to {\qcdad} include the large-$N$ volume independence and the large-$N$ orbifold/orientifold equivalences \cite{Lucini:2012gg} and the walking technicolor scenario \cite{Hill:2002ap}. A short review on {\qcdad} is available \cite{Shifman:2013yca}. 

Now let us consider {\qcdad} on $\RR^3 \times \SS^1$. As $\SS^1$ is not simply connected we need to specify boundary conditions for fields.  
In this paper we always assume the periodic boundary condition (PBC) for gauge fields. The boundary condition for fermions can be parametrized as
\begin{align}
  \Psi(\vec{x},x_4+L) = \ee^{i\phi} \Psi(\vec{x},x_4)\,,
  \label{eq:fbc}
\end{align}
where $L$ is the circumference of $\SS^1$. While $\phi=\pi$ (anti-periodic boundary condition, ABC) corresponds to 
thermal compactification, other choices are also of physical interest due to several reasons. To name but a few: 
\begin{itemize}
  \setlength{\itemsep}{-1.5mm}
  \item 
  PBC ($\phi=0$) is useful in supersymmetric gauge theories because it does not break SUSY \cite{Seiberg:1996nz}. 
  In SYM on $\RR^3\times \SS^1$ with PBC for fermions, various non-perturbative phenomena such as confinement and mass gap 
  generation have been shown analytically with semiclassical methods \cite{Davies:1999uw,Davies:2000nw,Poppitz:2012sw}. 
  \item 
  Twisted boundary condition \eqref{eq:fbc} offers a useful probe to QCD at finite temperature and density. 
  In $\SU(3)$ QCD with fundamental fermions, the partition function is periodic in $\phi$ with a period of $2\pi/3$ 
  (the Roberge-Weiss periodicity) and the dependence of observables on $\phi$ may distinguish the confined/deconfined phases \cite{Roberge:1986mm}. 
  It is also suggested that QCD at generic $\phi$ could help us investigate the phase 
  structure of finite-density QCD, because \eqref{eq:fbc} is equivalent to an \emph{imaginary} chemical potential, which causes no sign problem \cite{Alford:1998sd}. 
  Such a direction has been actively pursued with lattice simulations \cite{deForcrand:2002ci,D'Elia:2002gd,D'Elia:2004at,Chen:2004tb,deForcrand:2010he,Aarts:2013bla} 
  and chiral effective models \cite{Karbstein:2006er,Sakai:2008py,Sakai:2008um}. The twisted boundary condition is 
  also used to define the so-called dual quark condensate and the dressed Polyakov 
  loop \cite{Gattringer:2006ci,Bilgici:2008qy,Bilgici:2009jy,Bilgici:2009tx}. 
  \item 
  $\SU(N)$ {\qcdad} on $\RR^3\times \SS^1$ with $\phi=0$  
  exhibits gauge symmetry breaking $\SU(N)\to\U(1)^{N-1}$ \cite{Hosotani:1988bm}%
  \footnote{%
    This was shown by Hosotani for $\NfW\geq 2$ using a one-loop effective potential \cite{Hosotani:1988bm}. 
    For $\NfW=1$, by contrast, the bosonic and fermionic contributions 
    cancel to all orders in perturbation theory owing to supersymmetry and a non-perturbative treatment is required \cite{Poppitz:2012sw}. 
  }%
  , which is verified in lattice simulations \cite{Cossu:2009sq,Cossu:2013ora}. 
  This phenomenon, called Hosotani mechanism, is essential to the idea of 
  gauge-Higgs unification \cite{Manton:1979kb,Fairlie:1979at,Fairlie:1979zy,Hosotani:1983xw,Hosotani:1988bm}. 
  In perturbation theory, the low-energy spectrum after gauge symmetry breaking consists of free massless photons and fermions. 
  However this is not the end of the story: as first pointed out by \"Unsal, topological objects (monopole-instantons and their bound states called \emph{bions}) lead to mass gap and confinement in the small-$\SS^1$ semiclassical domain \cite{Unsal:2007vu,Unsal:2007jx}. 
  This can be viewed as a 4d generalization of Polyakov's treatment on the 3d Georgi-Glashow model \cite{Polyakov:1976fu} as well as  
  a non-SUSY generalization of preceding works on semiclassical confinement \cite{Seiberg:1994rs,Davies:1999uw,Davies:2000nw}. 
  This novel ``bion'' mechanism has been extended to fermions in other representations \cite{Shifman:2008ja,Poppitz:2009uq,Poppitz:2013zqa}. 
  For related works on instanton-monopoles and calorons, see Refs.~\cite{Lee:1997vp,Lee:1998bb,Kraan:1998pm,Kraan:1998sn,GarciaPerez:1999ux,Bruckmann:2003yq,
  Diakonov:2004jn,GarciaPerez:2006rt,Diakonov:2007nv,Unsal:2008ch,Diakonov:2008sg,GarciaPerez:2008gw,GarciaPerez:2009mg,Meisinger:2009ne,Diakonov:2009jq,
  Diakonov:2010qg,Anber:2011de,Ogilvie:2012is}%
  \footnote{Confusingly, the objects called `monopoles' or `monopole-instantons' in Refs.~\cite{Unsal:2007vu,Unsal:2007jx,Unsal:2008ch} are called `dyons' 
  in Refs.~\cite{Diakonov:2004jn,Diakonov:2007nv,Diakonov:2008sg,Diakonov:2009jq,Diakonov:2010qg}. 
  In this paper we will conform to the nomenclature of Refs.~\cite{Unsal:2007vu,Unsal:2007jx,Unsal:2008ch} for the reason in Ref.~\cite[footnote 6]{Poppitz:2012sw}.  
  Mutual relationship and (in)consistency of these works are discussed in Ref.~\cite[Section 4]{Poppitz:2012sw}.}. 
\end{itemize}
Other diverse applications of twisted boundary conditions (imaginary chemical potential) to QCD can be found 
in Refs.~\cite{Bedaque:2004kc,Mehen:2005fw,Damgaard:2005ys,Damgaard:2006pu,DeGrand:2006qb,Ozaki:2012ce}.

In this paper, motivated by these intriguing developments, 
we embark on the first systematic study of {\qcdad} on $\RR^3\times \SS^1$ with a twisted boundary condition \eqref{eq:fbc}. 
Here $\SS^1$ is a compactified \emph{spatial} direction.%
\footnote{\label{fn:com}At $\phi=\pi$ the system is equivalent to a thermal field theory at temperature $T=1/L$. 
Otherwise, we shall always view the compactified direction as a spatial direction, so all the phase transitions 
observed should be considered as zero-temperature quantum phase transitions. 
Note also that the VEV of a Polyakov loop that winds around a \emph{spatial} direction 
does not distinguish the physical confined/deconfined phases, 
because it is no longer related to the thermal free energy of excitations. Nevertheless, just for the sake of convenience, 
we will continue calling a center-symmetric (center-broken) phase a ``confined'' (``deconfined'') phase, respectively.}  
As the physics is periodic in $\phi$ modulo $2\pi$ it is sufficient to consider $0\leq \phi \leq \pi$.  
Despite a tremendous amount of literature on {\qcdad} with ABC $(\phi=\pi)$ and PBC ($\phi=0$), 
studies on intermediate $\phi$ are quite limited; we are only aware of partial perturbative analyses 
at one loop \cite{Hosotani:1988bm,Myers:2009df,Cossu:2013ora} and a qualitative discussion by Shuryak \cite{Shuryak:2008eq}. 
Intuitively, the effect of $\phi$ is anticipated to be negligible on large $\SS^1$. By contrast it has a dramatic impact 
on small $\SS^1$:  the system is in an Abelian confining phase at $\phi=0$ and in a hot deconfined plasma phase 
at $\phi=\pi$. How do these phases compete with each other at intermediate $\phi$? 
Is \"Unsal's bion mechanism of confinement still operative at nonzero $\phi$? We address these questions in this work. 

The realization of chiral symmetry is another important issue.  Due to asymptotic freedom, 
the coupling $g(L)$ is small at $L\ll \LQCD^{-1}$ and chiral symmetry will be restored there. 
What happens at intermediate $L$ is nontrivial. Based on the fact that the lowest Matsubara frequency 
of fermions obeying \eqref{eq:fbc} is $\phi/L$, one may naively expect that fermions become more relevant 
for smaller $\phi$ in the IR, leading to enhanced chiral symmetry breaking.   
It turns out, however, that this simple picture has to be modified because the Polyakov loop background 
shifts the net Matsubara frequency of fermions. Given that chiral dynamics is intertwined with center 
symmetry realization, it is essential for us to incorporate both at the same time for a correct understanding of the phase diagram. 

One more remark is in order regarding the fate of center symmetry at intermediate $L$. 
In {\qcdad} with $\NfW=1$ (SYM), the partition function with $\phi=0$ is nothing but the Witten index and is independent of $L$ \cite{Witten:1982df}, 
hence one expects no center-symmetry-changing transition at $0<L<\infty$. In {\qcdad} with $\NfW\geq 2$ and $\phi=0$, 
center symmetry is preserved at least for small $L$ (by one-loop effects) and large $L$ (by non-perturbative effects), but it is 
presently unclear if these domains are connected without any center-changing phase transition or not.  
If such a continuity were proven, it would mark significant progress towards a 
  first-principle understanding of IR renormalons in QCD, 
  as advocated in recent applications of resurgence theory to QFT \cite{Argyres:2012vv,Argyres:2012ka,Dunne:2012ae,Dunne:2012zk}.   
  However, lattice simulations \cite{Cossu:2009sq,Cossu:2013ora} as well as a model analysis \cite{Nishimura:2009me} 
  suggest that, in $\SU(3)$ {\qcdad} with light fermions at $\phi=0$, there appears a center-broken phase at intermediate $L$, 
  which calls into question the conjectured continuity between small $L$ and large $L$. 
  It also implies that the large-$N$ volume independence \cite{Kovtun:2007py,Unsal:2010qh},  
  whose premise is unbroken center symmetry, may fail at intermediate $L$. 
  Can we rescue the continuity by changing the gauge group to $\SU(2)$ or by considering a phase diagram on the $(L,~\phi)$-plane?  
  Answering this question constitutes part of the motivation of this work. 
  
To address a number of important issues raised above, we combine perturbation theory, the index theorem, semiclassical methods 
and a chiral effective model to investigate the phase structure of {\qcdad} as a function of $L$, $\phi,$ and the fermion mass $m$. 
Our main focus is on $N_c=2$ and $\NfW=2$, while the other cases are only briefly discussed.  On the phase diagram 
we find a number of quantum phase transition lines that separate phases with or without chiral and center symmetry breaking. 
Consistency of obtained results with our knowledge on the limiting cases ($\phi=0$, $\phi=\pi$ and $L\to 0$) is carefully examined, 
and the possibility of adiabatic continuity between the small-$L$ Abelian phase and the large-$L$ non-Abelian phase is scrutinized. 
All predictions in this work can be tested in future lattice simulations. 

This paper is organized as follows. In Section \ref{sec:centsym} we focus on the center symmetry realization. 
In Section \ref{sec:OL} we study center phase transitions in $\SU(2)$ and $\SU(3)$ gauge theories with
twisted boundary conditions by using a one-loop effective potential, which is reliable for $L\ll \LQCD^{-1}$. 
In Section \ref{sec:NP} we investigate the center phase structure by using a minimal non-perturbative gluon
effective potential that reproduces the deconfinement transition in pure Yang-Mills theory. 
By varying $m$ we interpolate 
between pure Yang-Mills theory ($m=\infty$) and massless {\qcdad} and delineate 
the evolution of the phase diagram.  
In Section \ref{sec:chisym} we use the the Polyakov--Nambu--Jona-Lasinio (PNJL) 
model \cite{Meisinger:1995ih,Fukushima:2003fw,Ratti:2005jh} to examine the influence of 
spontaneous chiral symmetry breaking on the phase structure of {\qcdad} in the chiral limit. 
First, we employ the high-temperature expansion to grasp qualitative features of the phase diagram and 
show that the Polyakov loop strongly affects chiral symmetry realization at $\phi\ne 0$. 
Then we move on to numerical analysis of the PNJL model, showing that the incorporation of 
chiral symmetry breaking modifies the phase diagram in a qualitative way. 
Among others, we find that 
the confining phase at $L\ll \LQCD^{-1}$ could be \emph{disconnected} from the confining phase at $L\gg \LQCD^{-1}$ 
on the plane spanned by $L^{-1}$ and $\phi$, depending on the parameters of the model.  
Section \ref{sec:SUM} is devoted to summary and discussions. 
In Appendix \ref{app:indextheorem}, we analyze zero modes of 
the Dirac operator on $\RR^3\times \SS^1$ for general $\phi$ on the basis of the index theorem 
in Refs.~\cite{Nye:2000eg,Poppitz:2008hr}.  
In Appendix \ref{app:modeltable} we review preceding studies on chiral effective models pertinent to 
the present work.

\section{Center symmetry realization}
\label{sec:centsym}
\subsection[Phase structure at small $\SS^1$]{\boldmath Phase structure at small $\SS^1$}
\label{sec:OL}

In this section we discuss center and gauge symmetry breaking at small $\SS^1$ using 
perturbation theory and semiclassical methods. In $\SU(N)$ {\qcdad}, the one-loop $\beta$ function 
of the running coupling is given by \cite{Peskin:1995ev}
\begin{align}
  \beta(g) = - \frac{g^3 N}{(4\pi)^2} \left(\frac{11}{3} - \frac{2}{3}\NfW \right) \,, 
  \label{eq:betafunc}
\end{align}
so the asymptotic freedom requires $\NfW\leq 5$. We will use $N$ and $N_c$ 
interchangeably to denote the number of colors in the rest of this paper. 

In considering a twisted boundary condition, one needs to carefully distinguish 
even $\NfW$ and odd $\NfW$. For simplicity let us begin with $\NfW=1$,   
with a boundary condition $\lambda(\vec{x},x_4+L)=\ee^{i\phi}\lambda(\vec{x},x_4)$ for the adjoint Weyl fermion $\lambda$. 
As discussed in Ref.~\cite[Section 5.2]{Poppitz:2008hr} (see also Refs.~\cite{Redlich:1984md,Niemi:1985ir}), 
fermions in this setup induce a Chern-Simons term in the three-dimensional effective theory on small $\SS^1$, whose 
coefficient depends on $\phi$. As is well known, the coefficient of the Chern-Simons term has to be properly quantized 
in order to maintain the invariance under large gauge transformations. It then follows that 
$\phi$ \emph{must be an integer multiple of} $\pi/N$ \cite{Poppitz:2008hr}; 
otherwise the gauge symmetry is spoiled and the theory is inconsistent.  

There is an intuitive way to see why only discrete values of $\phi$ are allowed. First, let us recall that 
the angle $\phi$ can be mapped to an imaginary chemical potential by field redefinition. 
Hence the partition function may be cast into the form 
\begin{align}
  Z(\phi) & = \Tr\!\big[ \!\ee^{-LH-i(\phi+\pi) N_F}\big]\,, 
  \label{eq:twistZ}
\end{align}
where $H$ is the Hamiltonian and $N_F$ is the fermion number. At $\phi=\pi$ it reduces to 
the thermal partition function $\Tr\!\big[\!\ee^{-LH}\big]$, as it should. Now what is problematic 
with \eqref{eq:twistZ} is that $N_F$ is not a conserved quantity: the $\U(1)$ fermion number symmetry 
at the classical level is broken to $\ZZ_{2N}$ by instantons. Since $N_F$ is 
conserved only modulo $2N$, the expression \eqref{eq:twistZ} is ill-defined, 
unless $(\phi+\pi) \cdot 2N$ is an integer multiple of $2\pi$. This leads to the quantization 
condition of Ref.~\cite{Poppitz:2008hr}. By the same token, for {\qcdad} with $\NfW\geq 2$ Weyl fermions 
of one chirality, the gauge invariance of the partition function requires that $\phi$ be an integer multiple of 
$\pi/(N \NfW)$. It is thus impossible to vary $\phi$ smoothly.  

However, when $\NfW$ is even, there is a way to avoid this pathology.   
By judiciously assigning $+\phi$ to $\NfW/2$ Weyl fermions and $-\phi$ to the other $\NfW/2$ Weyl fermions, 
the Chern-Simons term can be made to vanish after cancellation. This will happen automatically if we combine   
all the Weyl fermions into $\NfD(=\NfW/2)$ Dirac fermions, and impose a twisted boundary conditions \eqref{eq:fbc} 
on the latter. 
For this reason we will only consider Dirac fermions in the rest of this paper.%
\footnote{More generally, one should embed the $\U(1)$ boundary condition of fermions into a non-Abelian flavor symmetry 
that is anomaly free. This prescription works for odd $\NfW$ as well. We are grateful to M.~\"Unsal for pointing out this to us.} 

{\qcdad} is asymptotically free for $\NfW\leq 5$, hence $\NfD=1$ or $2$. 
Recent lattice simulations for $\SU(2)$ indicate that both $\NfD=1$ and $2$ may reside in the conformal window \cite{Catterall:2008qk,Hietanen:2008mr,Hietanen:2009az,DelDebbio:2009fd,DeGrand:2011qd,Athenodorou:2013eaa}, contrary to 
old simulations where chiral symmetry breaking was observed \cite{Kogut:1985xa,Kogut:1986jt}. As for $\SU(3)$, 
$\NfD=2$ is likely to be outside the conformal window \cite{Karsch:1998qj,Engels:2005te,Cossu:2009sq,Cossu:2013ora}, 
but there are uncertainties \cite{DeGrand:2013uha,DeGrand:2013yja}. 
The lattice data on the large-$N$ limit \cite{Azeyanagi:2010ne,Catterall:2010gx,Bringoltz:2011by,GonzalezArroyo:2012st,Gonzalez-Arroyo:2013bta,Gonzalez-Arroyo:2013dva} are not conclusive yet.  

Although our discussion in Section \ref{sec:OL} is unaffected by the absence or presence of IR conformality, 
Sections \ref{sec:NP} and \ref{sec:chisym} which concern  
non-perturbative aspects of {\qcdad} can be influenced by the location of the conformal window. 
Because the determination of the conformal window itself is not the purpose of this work, 
in Sections \ref{sec:NP} and \ref{sec:chisym} we will use effective models for specific values of $N$ and $\NfD$ 
where chiral symmetry breaking and confinement are simply \emph{assumed to occur} on $\RR^4$. After fixing the model 
parameters this way we will move on to the discussion of how the phase structure varies with $\phi$. Thus our model 
analysis should never be taken as constraining the range of conformal window. 

In the remainder of Section \ref{sec:OL}, we delineate the phase structure and dynamics at $L\ll\LQCD^{-1}$ 
using a perturbative potential and semiclassics. We will refer to a center-symmetric phase 
as a ``confined"  phase and to a center-broken phase as a ``deconfined" phase, with caveats of footnote \ref{fn:com} in mind.

\subsubsection[One-loop effective potential]{\boldmath One-loop effective potential}

In this section we summarize the one-loop effective potential for the Polyakov loop holonomy in $\SU(N_c)$ gauge theories 
on $\RR^3\times \SS^1$ with $\NfD$ adjoint Dirac fermions with a twisted boundary condition \eqref{eq:fbc}. Here we assemble 
relevant formulas for readers' convenience; full derivations can be found elsewhere \cite{Hosotani:1988bm,Myers:2009df,Cossu:2013ora}. 

To obtain the effective potential for the Polyakov loop holonomy, one needs to evaluate the functional determinant 
with a constant background field along the compact direction \cite{Gross:1980br,Weiss:1980rj} 
\begin{align}
  \label{eq:A4parametrization}
  \langle A_4 \rangle &=\frac{1}{L} \diag(q_1,q_2, \dots ,q_{N_c}) 
  \qquad \text{with} \quad  \sum_{k=1}^{N_c}q_k=0\,. 
\end{align}
The holonomy $\Omega$ and the normalized traced Polyakov loop  
in the fundamental representation $\PF$ are defined by
\begin{align}
  \Omega \equiv \mathcal{P}\exp\left(i\oint dx_4\, A_4\right) \qquad 
  \text{and} \qquad 
  \PF \equiv \frac{1}{N_c} \Tr \Omega \,,
\end{align}
where $\mathcal{P}$ denotes path-ordering. 
The gluon$+$ghost contribution to the one-loop effective potential ${\cal V}_{\YM}$ reads 
\begin{align}
  {\cal V}_{\YM}(N_{c}, L; \{q\})
  & = - \frac{2}{\pi^2L^4} \sum_{n=1}^{\infty} \frac{1}{n^4} \Tr_{\adj}(\Omega^n)
  \\
  &= - \frac{2}{\pi^2 L^4} \sum_{n=1}^{\infty} \frac{1}{n^4} \sum_{i,j=1}^{N_{c}} 
     \Big( 1 - \frac{1}{N_{c}} \delta_{ij} \Big) \cos( n q_{ij}) \,, 
  \label{gc}
\end{align}
with $q_{ij} \equiv  q_i - q_j $. 

The adjoint fermion contribution consists of a ``zero-temperature'' part ${\cal V}_\chi$ 
and a ``thermal'' part ${\cal V}_{\adj}$.%
\footnote{We once again emphasize that this terminology is used only for 
convenience; $\SS^1$ in this work is considered as a compact spatial direction.} 
As ${\cal V}_\chi$ has no $\{q\}$-dependence it can be dropped 
throughout Section \ref{sec:centsym}. (It will be retrieved later in Section \ref{sec:chisym} where 
we discuss spontaneous chiral symmetry breaking.) ${\cal V}_{\adj}$ is given by
\begin{align}
  & {\cal V}_{\adj}(N_{c}, \NfD, L, m, \phi; \{q\}) 
  \notag
  \\
  & = \frac{\NfD m^2}{\pi^{2} L^{2}} \sum_{n=1}^{\infty} \frac{K_2(nLm)}{n^2}
  \big[ \ee^{in\phi} \Tr_{\adj}(\Omega^n) + \ee^{-in\phi} \Tr_{\adj}(\Omega^{\dagger n}) \big]
  \label{preadjc}
  \\
  & =   
  \frac{ 2 \NfD m^2}{\pi^{2} L^{2}} \sum_{n=1}^{\infty} 
  \frac{K_2 ( n L m )}{n^2} \sum_{i,j=1}^{N_{c}} \Big(  1 - \frac{1}{N_{c}} \delta_{ij} \Big)
  \cos \big(n (q_{ij} + \phi)\big) \,,
  \label{adjc}
\end{align}
where $m$ is the fermion mass,  $0\leq\phi\leq\pi$ is the boundary twist, 
and $K_\nu(x)$ is the modified Bessel function of the second kind. As a small consistency check, 
we verify that ${\cal V}_\YM + {\cal V}_{\adj}$ vanishes exactly for $\phi=0$, $\NfD=1/2$ and $m\to 0$, 
as expected from supersymmetry.

\subsubsection[Phase diagram for $(N_c\,,\NfD)=(2,1)$, $(3,1)$ and $(3,2)$]
{\boldmath Phase diagram for $(N_c\,,\NfD)=(2,1)$, $(3,1)$ and $(3,2)$}
\label{sec:pd_ncnf}

\begin{figure}[t]
  \centerline{
    \includegraphics[width=0.97\textwidth]{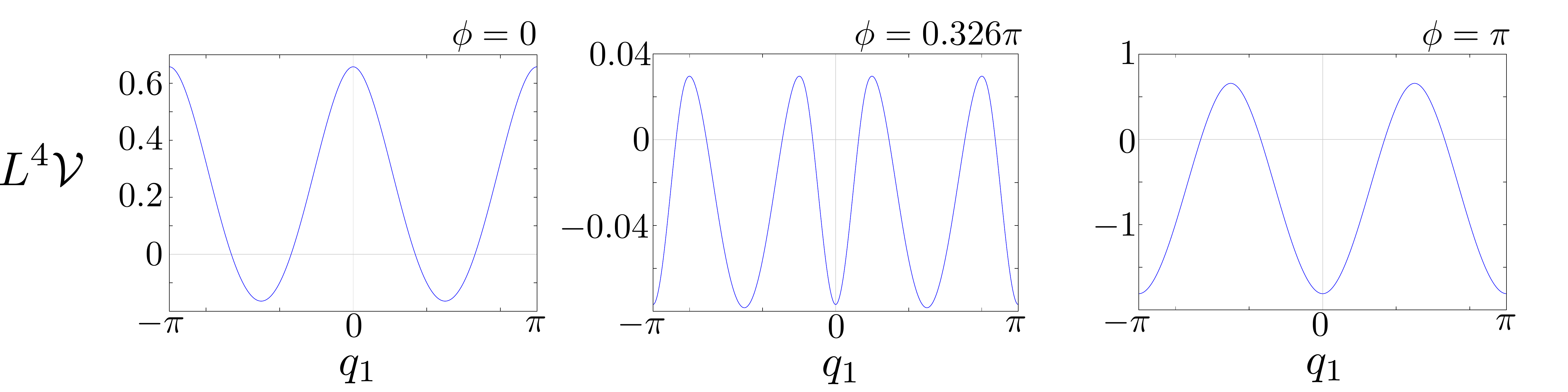}
  }
  \vspace{-.3\baselineskip}
  \caption{\label{Nc2Nf1}%
  ${\cal V}(N_{c}=2, \NfD=1)$ 
  in \eqref{VNc2Nf1} at $m=0$ for three values of $\phi$. 
  A first-order phase transition is seen to occur at $\phi\simeq 0.326\pi$.}
\end{figure}

Let us begin with $\SU(2)$ {\qcdad} with $\NfD=1$. 
The one-loop effective potential in this case is given by
\begin{equation}
  {\cal V}(N_{c}=2, \NfD=1) = {\cal V}_{\YM}(N_{c}=2, L; \{q\}) + 
  {\cal V}_{\adj}(N_{c}=2, \NfD=1, L, m, \phi; \{q\})\,,
  \label{VNc2Nf1}
\end{equation}
with the condition $q_{1}+q_{2}=0$. 
The potential $L^{4}\mathcal{V}(N_{c}=2, \NfD=1)$ in the massless limit $(m=0)$
is plotted in Figure~\ref{Nc2Nf1} for $-\pi \leq q_1 \leq \pi$. 
For $0\leq \phi <0.326\pi$, the global minima are located at $(q_1,q_2)=(\pm\pi/2, \mp \pi/2)$, 
thus the VEV of the Polyakov loop is 
\begin{equation}
  \langle \PF \rangle = \frac{1}{2} \Tr \begin{pmatrix} \ee^{\pm i\pi/2} & 0 \\ 0 & \ee^{\mp i\pi/2} \end{pmatrix} =0\,,
\end{equation}
which indicates that the $\ZZ_{2}$ center symmetry is intact in this phase.
It is also notable that the $\SU(2)$ gauge symmetry is broken to $\U(1)$ since $q_{1}\not=q_{2}$. 
At $\phi\sim0.326\pi$ there is a first-order phase transition
to the vacuum with $(q_1,q_2)=(0,0)$ and $(\pm\pi,\mp\pi)$, for which 
\begin{equation}
  \langle \PF \rangle = \frac{1}{2}\Tr \begin{pmatrix} \pm 1 & 0 \\ 0 & \pm 1 \end{pmatrix} =\pm 1\,. 
\end{equation}
This phase breaks $\ZZ_{2}$ spontaneously but preserves the $\SU(2)$ gauge symmetry.  
In summary, the phase structure for $m=0$ is given as follows: 
\begin{equation}
  \label{eq:table_2_1}
\begin{tabular}{|c||c|c|c|}
  \hline
  $(N_c,\NfD)=(2,1)$ & Phase & $\langle \PF \rangle$ & Gauge sym.
  \\\hline\hline 
  $0\leq \phi <0.326\pi$ &  Confined & $0$ & $\U(1)$ 
  \\\hline 
  $0.326\pi <\phi \leq \pi$ &  Deconfined & $\pm1$ & $\SU(2)$ 
  \\\hline 
\end{tabular}
\end{equation}
Our result is consistent with Ref.~\cite[Theorem 5]{Hosotani:1988bm}.

\begin{figure}[t]
  \centerline{
    \includegraphics[width=0.4\textwidth]{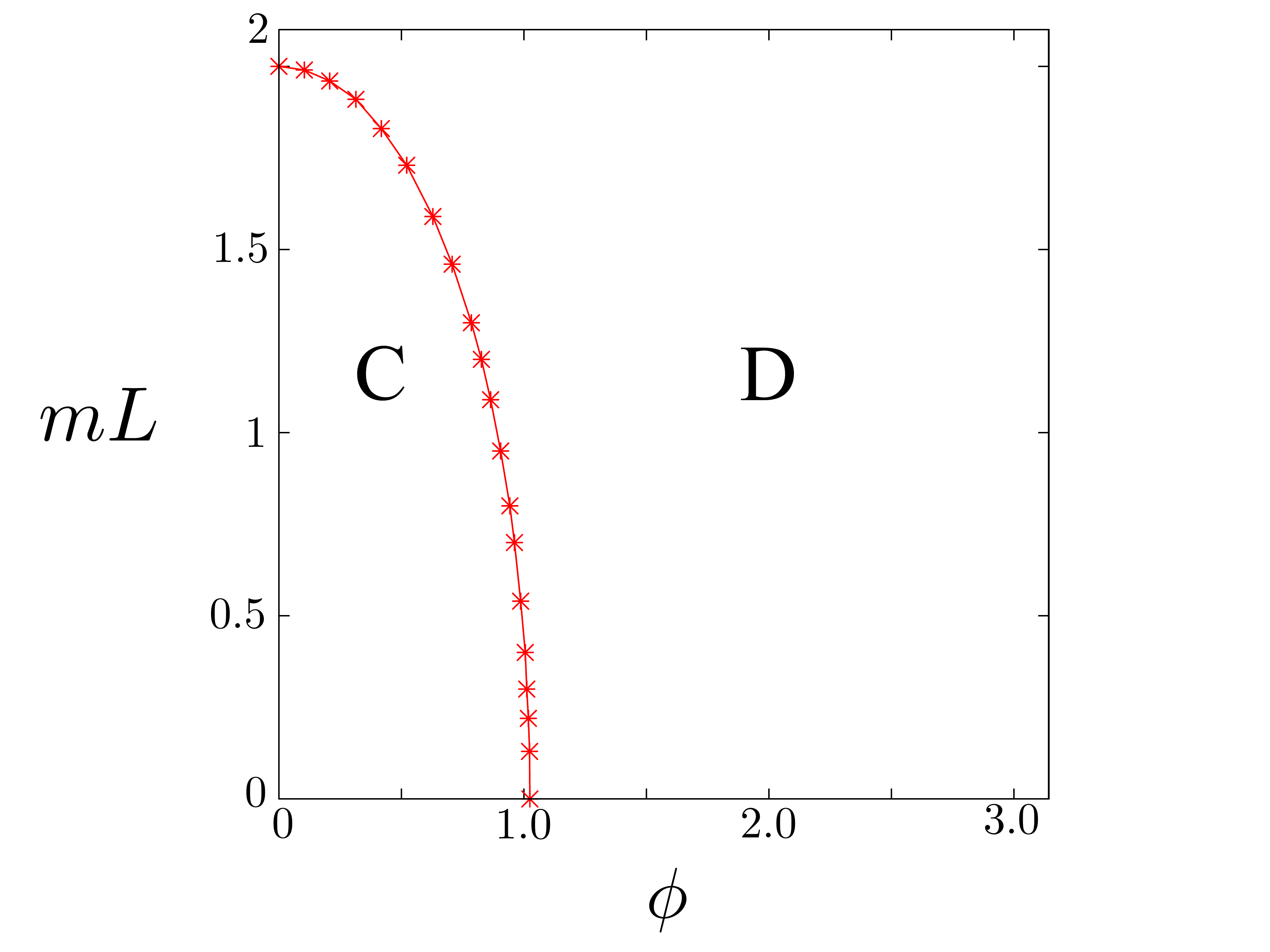}\quad
    \includegraphics[width=0.4\textwidth]{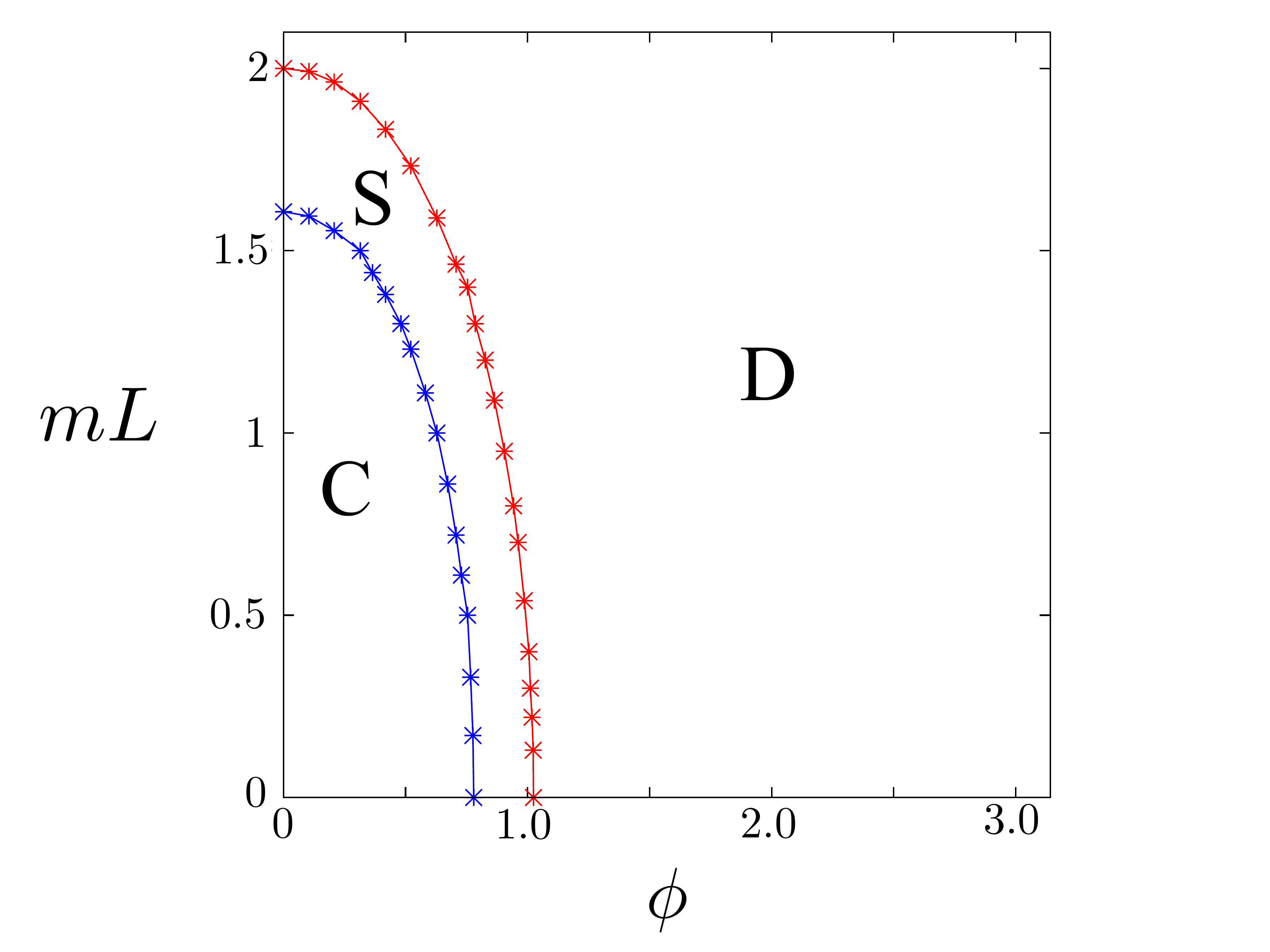}
    \put(-282,132){$N_c=2$}
    \put(-98,132){$N_c=3$}
  }
  \vspace{-.3\baselineskip}
  \caption{\label{PNc2Nf1}  
    The $mL$--$\phi$ phase diagram (\textbf{Left:} $N_{c}=2$, $\NfD=1$. \textbf{Right:} $N_{c}=3$, $\NfD=1$).  
    The symbols C, D and S refer to the confining phase, the deconfined phase and the split phase, respectively. 
    In both figures, the transitions are first order. 
  }
\end{figure}

For $m\not=0$, the shape of the potential \eqref{VNc2Nf1} depends on $\phi$ and $Lm$. 
As shown in Figure~\ref{PNc2Nf1} (left), the confining phase is favored for small $Lm$ and small $\phi$, 
whereas the deconfined phase is favored throughout the rest of the phase diagram. 
This can be qualitatively understood from the fact that the adjoint fermions decouple from the low-energy physics 
when either the current mass or the lowest Matsubara frequency $\sim \phi/L$ is large, leaving gluons that 
favor the deconfined phase at small $\SS^1$. 

\begin{figure}[t]
  \centerline{ 
    \includegraphics[width=0.85\textwidth]{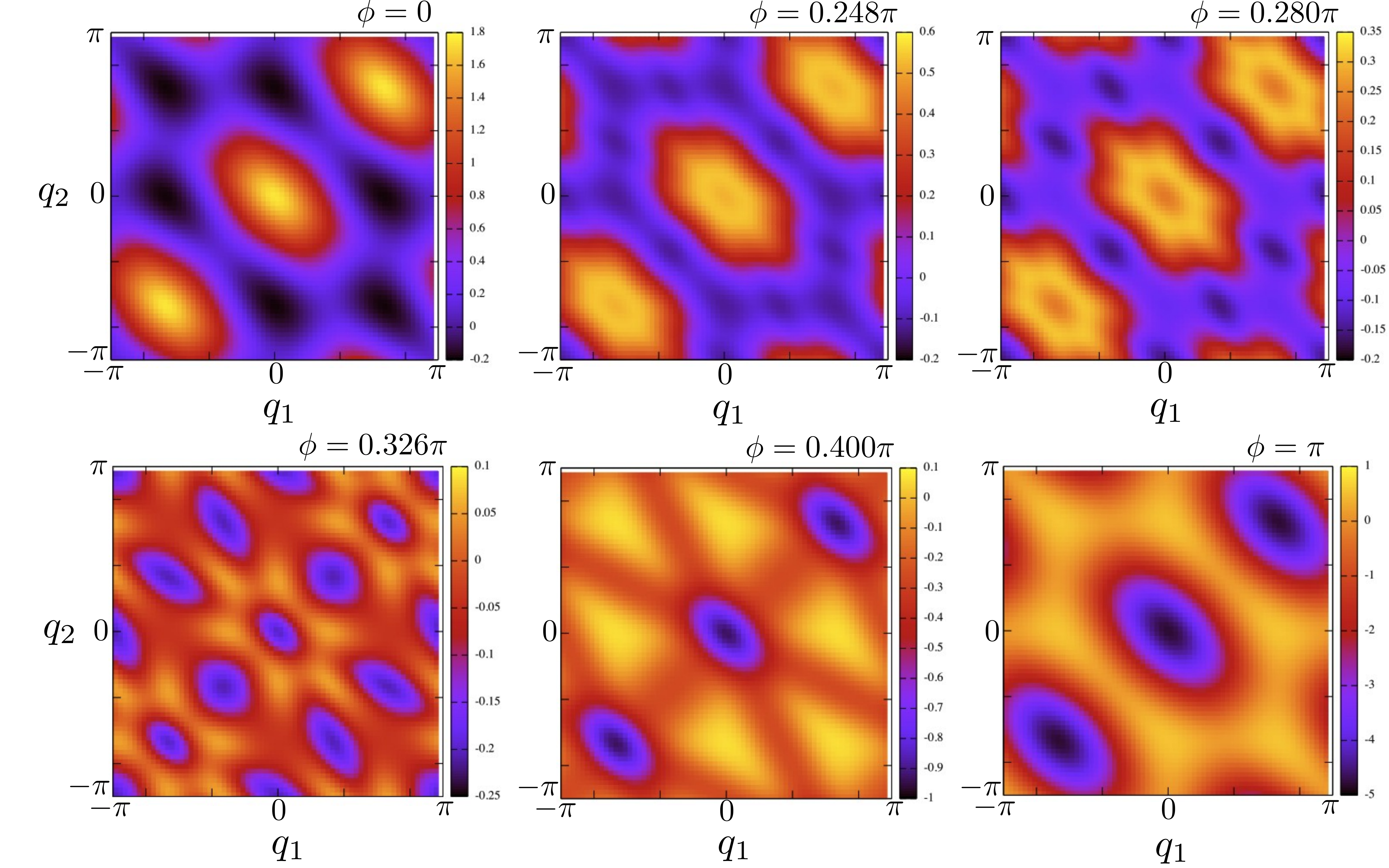}
  }
  \vspace{-.3\baselineskip}
  \caption{\label{Nc3Nf1}%
  Contour plots of ${\cal V}(N_{c}=3, \NfD=1)$ 
  in \eqref{VNc3Nf1} at $m=0$ and $\phi \in \{0,~0.248\pi,~0.280\pi,$ $0.326\pi,~0.400\pi,~\pi\}$. 
  Phase transitions occur at $\phi\simeq 0.248\pi$ and $\phi\simeq 0.326\pi$, from the confining 
  to the split phase and then to the deconfined phase, respectively. 
  }
\end{figure}

Next, we consider $\SU(3)$ {\qcdad} with $\NfD=1$. 
The one-loop effective potential for this case is given by
\begin{equation}
  {\cal V}(N_{c}=3, \NfD=1) = {\cal V}_{\YM}(N_{c}=3, L; \{q\}) + 
  {\cal V}_{\adj}(N_{c}=3, \NfD=1, L, m, \phi; \{q\})\,,
  \label{VNc3Nf1}
\end{equation}
with the condition $q_{1}+q_{2}+q_{3}=0$. 
For massless case $m=0$, the potential $L^{4}\mathcal{V}(N_{c}=3, \NfD=1)$ is depicted in Figure~\ref{Nc3Nf1} 
as a function of $-\pi\leq q_1,\,q_2\leq \pi$ and $\phi$. 
For $0\leq \phi <0.248\pi$, the global minima are given by the six permutations of 
$(q_{1},q_{2}, q_{3})=(0,2\pi/3,-2\pi/3)$, and 
\begin{equation}
  \langle \PF \rangle = \frac{1}{3} \Tr\begin{pmatrix} 1 &0&0\\0&\ee^{2i\pi/3}&0\\0&0&\ee^{-2i\pi/3} \end{pmatrix} =0\,. 
\end{equation}
This phase is $\ZZ_3$-symmetric but breaks the $\SU(3)$ gauge symmetry down to $\U(1)\times \U(1)$.  
At $\phi\sim0.248\pi$ there is a first-order phase transition
to the vacuum with $(q_{1},q_{2}, q_{3})=(0,\pm\pi,\mp\pi)$, $(\pm 2\pi/3, \mp\pi/3, \mp\pi/3)$ and their permutations, for which 
\begin{equation}
  \langle \PF \rangle = - \frac{1}{3},~\frac{\ee^{i \pi/3}}{3}, ~~\text{and}~ ~\frac{\ee^{- i \pi/3}}{3} \,. 
\end{equation}
This is the so-called ``split phase'' \cite{Myers:2009df,Cossu:2009sq} in which 
$\SU(3)$ gauge symmetry is broken to $\SU(2)\times \U(1)$ since only two of the eigenvalues of the holonomy  
are degenerate. The $\ZZ_{3}$ symmetry is also broken by $\langle \PF \rangle\ne 0$.  
Finally, at $\phi\sim0.326\pi$ there is another first-order phase transition 
to the vacuum with $(q_{1},q_{2}, q_{3}) = (0,0,0)$ and $(\pm 2\pi/3,\pm 2\pi/3,\pm 2\pi/3)$, with 
the Polyakov loop VEV 
\begin{equation}
  \langle \PF \rangle = 1,~ \ee^{2i\pi/3},~~ \text{and} ~~ \ee^{-2i\pi/3}\,. 
\end{equation}
This is the usual deconfined phase. The overall center phase structure 
for $m=0$ is summarized below. 
We note that a similar phase structure has been observed in pure Yang-Mills theory 
with deformation \cite{Myers:2007vc}. 
\begin{equation}
\begin{tabular}{|c||c|c|c|}
  \hline 
  $(N_c,\NfD)=(3,1)$ & Phase & $\langle \PF \rangle$ & Gauge sym.
  \\\hline \hline 
  $0\leq \phi <0.248\pi$ &  Confined & $0$ & $\U(1)\times \U(1)$
  \\\hline 
  $0.248\pi < \phi <0.326\pi$ & Split  & $-1/3,\ee^{\pm i \pi/3}/3$ & $\SU(2)\times \U(1)$ 
  \\\hline 
  $0.326\pi < \phi \leq \pi$ & Deconfined & $1,\ee^{\pm i 2\pi/3}$ & $\SU(3)$ 
  \\\hline
\end{tabular}
\end{equation}

For $m \not=0$, the potential \eqref{VNc3Nf1} depends on $Lm$. 
The phase structure, depicted in Figure~\ref{PNc2Nf1} (right), is analogous to 
the $\SU(2)$ case as a whole. It is intriguing that the confined and the deconfined phases are always separated 
by the split phase, although no such intermediate phase appears around the finite-temperature transition in $\qcdad$. 
As $N_c$ is increased there will be even more exotic phases with broken gauge symmetry, between the confined 
phase at $\phi=0$ and the deconfined phase at $\phi=\pi$. 
\\\ 

Finally let us briefly discuss the case with $N_c=3$ and $\NfD=2$, for which we only show 
the result for $m=0$ below.  This result is consistent with Ref.~\cite[eq.~(4.11)]{Cossu:2013ora}. 
\begin{equation}
\begin{tabular}{|c||c|c|c|}
  \hline 
  $(N_c,\NfD)=(3,2)$ & Phase & $\langle \PF \rangle$ & Gauge sym.
  \\ \hline \hline 
  $0\leq \phi <0.319\pi$ &  Confined & $0$ & $\U(1)\times \U(1)$ 
  \\\hline 
  $0.319\pi < \phi <0.416\pi$ & Split  & $-1/3,\ee^{\pm i \pi/3}/3$ & $\SU(2)\times \U(1)$ 
  \\\hline 
  $0.416\pi < \phi \leq \pi$ & Deconfined & $1,\ee^{\pm i 2\pi/3}$ & $\SU(3)$
  \\\hline
\end{tabular}
\end{equation}
While the overall phase structure is the same as 
the previous $\NfD=1$ case, the confined (deconfined) phase has enlarged (shrunk), as there are 
more fermions that tend to favor the confining phase at small $\phi$. 

We here comment on the validity and limitation of the perturbative one-loop effective potential. 
The method may be used to calculate the leading order contribution to the free energy and 
determine the phase structure at least in the weak-coupling regime with $L\ll \LQCD^{-1}$.  
However perturbation theory is not only blind to the strong-coupling physics such as chiral symmetry breaking and 
confinement on large $L$ but also misses a plethora of non-perturbative phenomena 
induced by topological objects at small $\phi$ for $L\ll \LQCD^{-1}$. We will discuss the latter in the next subsection.

\subsubsection{Mass gap and confinement from semiclassics}
\label{eq:semicla}

Before considering {\qcdad} at general $\phi$, let us recapitulate the semiclassical physics of {\qcdad} at $\phi=0$ following Refs.~\cite{Unsal:2007vu,Unsal:2007jx}. 
In this subsection we always assume that the bare mass of fermions is zero.  

First of all, in QCD at high temperature $T\gg \LQCD$, 
the coupling goes small due to asymptotic freedom and the semiclassical instanton gas approximation is justified \cite{Gross:1980br}.   
On the other hand, the magnetic sector remains strongly coupled at any $T$: the fermions with ABC decouple from the dynamics at high $T$ and the theory reduces 
to the three-dimensional Yang-Mills theory, which generates non-perturbative mass gap and exhibits an area law of confinement for the spatial Wilson loop, 
precluding a naive application of perturbation theory. 

The situation is drastically different if we go to {\qcdad} with PBC at small $\SS^1$. The major differences, as compared to thermal QCD, are as follows.%
\footnote{See Ref.~\cite{Anber:2011de} for a nice review on this subject.}  
\begin{itemize}
  \item 
  The adjoint fermions with PBC stabilizes a confining phase at small $L$, as shown in Section \ref{sec:pd_ncnf}. They never decouple in the IR 
  since they have a zero mode along $\SS^1$. 
  \item 
  The $\SU(N)$ gauge symmetry is broken to $\U(1)^{N-1}$ and as a result of which, the off-diagonal components of the gauge fields and fermions 
  acquire a large mass $\sim 1/L$ via the Higgs mechanism. For $\NfW\geq 2$, the electric gluons $A_4^a$ also acquire a mass $\sim g/L$ from the one-loop 
  effective potential.  
  \item 
  The perturbative low-energy effective theory on $\RR^3$ consists of massless photons and fermions, 
  which are neutral under $\U(1)^{N-1}$ and hence non-interacting.  
  Since charged particles are absent below the scale $\sim 1/L$, the running coupling $g(\mu)$ ceases to run at $\mu\sim 1/L$; 
  $g(1/L)$ is small provided $1/L\gg \LQCD$%
  \footnote{Here we assume $N = {\cal O}(1)$. In general, a weak-coupling analysis is valid if $NL\LQCD\ll 1$ \cite{Unsal:2010qh}.} 
  and validates the semiclassical expansion. 
  \item 
  In the background of $\ZZ_N$-symmetric Polyakov loop, instantons split into $N-1$ BPS monopoles and 
  one KK monopole \cite{Lee:1998bb,Kraan:1998sn,Kraan:1998pm}, 
  which interact with each other via a 3d Coulomb potential. They are more relevant than instantons 
  due to their smaller action $\Big(S_0=\frac{8\pi^2}{g^2N}<\frac{8\pi^2}{g^2} \Big)$. 
  \item 
  Monopoles alone do {\it not} contribute to the bosonic potential, because each monopole is accompanied by 2 fermionic zero modes (per flavor) 
  in accordance with the Index theorem. (This is to be contrasted with Polyakov's 3d $\U(1)$ model \cite{Polyakov:1976fu} where the monopoles do generate a mass gap.) 
  Instead, it is the bound states of BPS and $\overline{\rm KK}$ monopoles called \emph{magnetic bions}%
  \footnote{This is a simplified statement valid for $\SU(2)$. For general $\SU(N)$, precisely speaking, a ``magnetic bion'' is a bound state of 
  a monopole of type $i$ and an anti-monopole of type $i+1$ for $i=1,\dots,N$ \cite{Unsal:2007jx}.}, that generate a mass gap  
  of order $L^{-1} \ee^{-S_0}$ for photons \cite{Unsal:2007vu,Unsal:2007jx}.  
  They are topologically neutral and carry magnetic charge 2. It is the attractive interaction due to massless fermion exchange that overcomes 
  the Coulomb repulsion between BPS and {$\overline{\rm KK}$} monopoles, leading to the formation of molecules.    
\end{itemize}
Now we are prepared to ask what occurs when $\phi\ne 0$. As shown in the last subsection, the confining phase 
with $\langle \PF \rangle=0$ is sustained for $0\leq \phi<\phi_c$ for some critical $\phi_c$ 
which depends on $N$ and $\NfD$. In this phase, the Polyakov-loop holonomy is 
$\ZZ_N$-symmetric and breaks the $\SU(N)$ gauge symmetry down to $\U(1)^{N-1}$, in much the same way as at $\phi=0$.  
But a new feature shows up in the three-dimensional perturbative effective theory%
\footnote{In this subsection we assume $0\leq \phi< \phi_c<\pi$. Formulas valid for generic $\phi$ can be easily obtained from those in this subsection 
by replacing $\phi$ with the lowest Matsubara frequency, i.e., $\min\{|2\pi n + \phi| \ \big|\, n \in \ZZ \}$. }: 
\begin{align}
  S_{\rm eff} = \int_{\RR^3} d^3x~\frac{L}{g^2} \sum_{\ell=1}^{N-1} \Bigg[ \frac{1}{4}{F^{(\ell)}_{ij}}^2 + 
  \sum_{f=1}^{\NfD} \bar\Psi^{(\ell)}_f\left( \gamma_i \der_i + i\frac{\phi}{L}\gamma_4 \right)\Psi^{(\ell)}_f \Bigg]\,,
\end{align}
with $i,j=1,2,3$ denoting the three spatial directions. Here $\Psi^{(\ell)}_f$ are adjoint Dirac spinors with $\ell$ labeling the diagonal components, and 
the term $\propto\phi/L\equiv m_\psi$ originates from the twisted boundary condition.%
\footnote{This term was used in Ref.~\cite{Shifman:2009tp} as an IR regulator for fermions.} The latter is often called ``real mass'' 
to distinguish it from the complex Dirac mass. For this effective theory to be valid, the cutoff scale $\mu$ 
must satisfy $m_\psi \ll\mu\ll g/L$, which necessitates $\phi\ll g$. If $\phi={\cal O}(1)$, fermions are integrated out as well, and 
one ends up with a theory of free massless photons.  So much for the perturbation theory. 

Now we turn to the non-perturbative dynamics. Associated with the gauge symmetry breaking, 
there appear $N$ types of monopoles ($N-1$ BPS and one KK). 
On $\RR^3$ each BPS monopole carries $2\NfW$ adjoint zero modes ($2$ per flavor), as dictated by the Callias index theorem. 
On $\RR^3\times \SS^1$ the index generally depends on the boundary condition $\phi$, but as will be shown in Appendix \ref{app:indextheorem}, 
the number of adjoint zero modes on each monopole is \emph{independent of} $\phi$ and equals $2\NfW$ for the particular case of a 
$\ZZ_N$-symmetric Polyakov-loop background. 
(See the left panel of Figure \ref{fg:N2N3index} in Appendix \ref{app:indextheorem}, 
where the index at $q=\pi/2$ is equal to 2 for all values of $\phi$.)   
This implies that, for $0\leq \phi<\phi_c$, 
monopoles interact with each other via a sum of a Coulomb potential and an attractive potential induced by fermion zero-mode exchange.  
This picture is essentially the same as at $\phi=0$. 

To see a new phenomenon specific to $\phi\ne 0$, let us turn to the expression for the bion amplitude \cite{Unsal:2007jx,Anber:2011de,Argyres:2012ka}
\begin{align}
  Z_{\rm bion}(g) & \sim \frac{1}{g^8}\exp\left(- \frac{2}{N} \frac{8\pi^2}{g^2} (1+cg) \right)
  \int d^3x \int d^3y ~ \exp\left(-\frac{4\pi L}{g^2|\vec{x}-\vec{y}|}\right) \big[ S_F(\vec{x}-\vec{y}) \big]^{2\NfW} \,,
  \label{eq:bioz}
\end{align}
where the factor $1/g^8$ is associated with the Jacobian for collective coordinates, the first exponential is the weight for two monopoles 
(with a correction $\propto cg$, $c={\cal O}(1)$, arising from the non-BPS nature of monopoles \cite{Kirkman:1981ck}), 
the second exponential is the Coulomb interaction between monopoles located at $\vec{x}$ and $\vec{y}$, and the final factor 
is the fermion zero-mode exchange interaction, which can be extracted from 
the correlator of 't Hooft vertices \cite{Unsal:2007jx}.%
\footnote{In \eqref{eq:bioz} we are a bit cavalier about spinor indices but this is inessential to the ensuing discussion.}    
The free fermion propagator $S_F$ is defined as  
\begin{align}
  S_F(\vec{r}) & \equiv \left. \int \! \frac{d^3p}{(2\pi)^3} \frac{\ee^{i \vec{p}\cdot \vec{r}}}{\sigma_i p_i + im_\psi}\right|_{m_\psi=\phi/L}
  \\
  & = \frac{i}{4\pi}\ee^{-m_\psi r} \left\{
    \left(\frac{1}{r^3}+\frac{m_\psi}{r^2}\right)\sigma_i x_i - \frac{m_\psi}{r}\1 
  \right\}\,, 
\end{align}
where the higher Matsubara frequencies are not relevant to our argument and are omitted above.  
Due to the real mass $m_\psi$, the fermion propagator has an exponential fall-off at large $r$.   
At this point we emphasize the difference between the real mass $m_\psi$ and the 
complex Dirac and Majorana masses; the former does not yield a disconnected 
piece for the 't Hooft vertex correlator, whereas the latter soak up zero modes of 
monopoles and allow unpaired monopoles to contribute to the bosonic potential \cite{Poppitz:2012sw}. 

Making a crude approximation $S_F(\vec{r})\sim \ee^{-m_\psi r}$, one finds  
\begin{align}
  Z_{\rm bion} & \sim \frac{1}{g^8}\exp\left(- \frac{2}{N} \frac{8\pi^2}{g^2}(1+cg) \right) \int_{r_{\min}}^\infty dr~r^2~\exp\left(-V_{\rm eff}(r)\right)\,, 
  \label{eq:Zbion}
\end{align}
where $r_{\min}\sim L$ is a cutoff%
\footnote{The integral is convergent even if $r_{\rm min}$ is set to zero; however, the interaction between monopoles 
at short distance cannot be approximated by a Coulomb potential and the argument based on \eqref{eq:veff}  
loses its meaning.} (irrelevant for the following discussion) and  
\begin{align}
  V_{\rm eff}(r) = \frac{4\pi L}{g^2 r} + 2 \NfW m_\psi r  \,. 
  \label{eq:veff}
\end{align}
The minimum of $V_{\rm eff}(r)$ is located at $r=r_{\rm b}$ with 
\begin{align}
  r_{\rm b} & \equiv \sqrt{ \frac{2\pi L}{ g^2\NfW m_\psi} } 
  = \sqrt{\frac{2\pi}{\NfW \phi}}\,\frac{L}{g}\,.
  \label{eq:bionsize}
\end{align}
This gives a rough size of bions. 
It is noteworthy that $r_{\rm b}$ is {\it far smaller} than the bion size $\sim L/g^2$ 
at $\phi=0$ \cite{Anber:2011de,Argyres:2012ka}. This is because the linear confining potential $\sim m_\psi r$ in $V_{\rm eff}(r)$ 
binds monopoles more strongly than the logarithmic potential $\sim \log(m_{\psi}r)$ that acts at $\phi=0$.  
This mechanism is reminiscent of the instanton--anti-instanton molecules in QCD at high temperature, 
where the zero-mode exchange induces a linear potential between instantons and anti-instantons \cite{Khoze:1990nt,Schafer:1994nv}. 

Now let us perform three consistency checks: 
\begin{itemize}
  \item 
  $r_{\rm b}\gg L$ for $\phi = {\cal O}(1)$, so the usage of the Coulomb potential is valid. 
  \item 
  $\log(m_{\psi}r_{\rm b}) \ll m_\psi r_{\rm b}$ implies that the linear potential 
  is indeed the dominant part of the interaction of monopoles on the scale $r\approx r_{\rm b}$, 
  which justifies our approximation. 
  \item 
  In the above treatment the electric interaction of monopoles has been ignored. This simplification is innocuous 
  at $\phi=0$ because the distance between monopoles $\sim L/g^2$ is much longer than 
  the inverse of the gap $\sim g/L$ of $A_4$. By contrast, at $\phi\ne 0$, $r_{\rm b}^{-1}$ is 
  comparable to the gap of $A_4$ and, precisely speaking, the $A_4$-exchange interaction could be 
  as important as the other interactions.  In Ref.~\cite[eq.~(5.10)]{Argyres:2012ka} a full effective 
  potential between monopoles incorporating the electric interaction has been worked out, where 
  $A_4$-exchange modifies the ${\cal O}(1)$ prefactor of the Coulomb interaction, but never changes an overall    
  form of the potential. On the basis of this observation, we conclude that 
  the electric interaction would not invalidate our argument here, even though it may well change 
  the ${\cal O}(1)$ prefactor of $L/g$ in \eqref{eq:bionsize}. 
\end{itemize}
In order to evaluate the integral \eqref{eq:Zbion} we switch to the dimensionless variable $x\equiv r/r_{\rm b}$, 
finding 
\begin{align}
  Z_{\rm bion} \sim \frac{1}{g^8}\exp\left(- \frac{2}{N} \frac{8\pi^2}{g^2}(1+cg) \right) (r_{\rm b})^3 \! 
  \int_{x_{\rm min}}^{\infty}\!\!\! dx~x^2 \exp\Bigg(\! -\frac{2\sqrt{2\pi \NfW \phi}}{g}\Big(x+\frac{1}{x}\Big) \Bigg) \,.
\end{align}
The integral is dominated by the contribution from $x\approx 1$. Performing a gaussian approximation 
around $x=1$, which is accurate for $g\ll 1$, we finally obtain
\begin{align}
  Z_{\rm bion} \sim \frac{1}{\phi^{7/4}g^{21/2}}\exp\Bigg(\! - \frac{2}{N} \frac{8\pi^2}{g^2}(1+cg) - \frac{4\sqrt{2\pi\NfW\phi}}{g}~ \Bigg)\,. 
  \label{eq:zbion}
\end{align} 
This result is valid regardless of whether $\NfW$ is inside the conformal window or not. As is evident from \eqref{eq:zbion}, the emergence of the factor $\sim 1/g$ in the exponent shows that the bions at $\phi\ne 0$ are {\it exponentially 
suppressed} compared to the case at $\phi=0$.%
\footnote{The contribution due to $\phi\ne 0$ in the exponent is parametrically of the same order 
as the correction from $cg$; this is the reason why we have kept the latter in all equations so far. }  
An intuitive explanation for this is that, owing to the strong binding force 
between monopoles, they are brought closer together and feel the Coulomb repulsion more severely than at $\phi=0$, which makes 
bions energetically costlier. A natural question to ask is: when does this mechanism set in if $\phi$ is smoothly 
increased from $0$? The linear potential due to the exchange of massive fermions will begin to matter only when 
$m_\psi^{-1}=L/\phi$ becomes comparable to the bion size $L/g^2$. This suggests that the bion suppression sets in for $\phi\gtrsim g^2$.  

Alongside the dilution of bions, the bion-induced non-perturbative quantities such as the mass gap and the spatial string tension are  
also expected to be exponentially suppressed compared to $\phi=0$. To see this, let us compute the mass gap ${\cal M}$ of 
photons following Refs.~\cite{Polyakov:1976fu,Anber:2011de} for $N=2$. 
Using the two-loop running coupling 
\begin{align}
  \frac{g(L)^2}{4\pi}= \frac{2\pi}{\beta_0 \log \frac{1}{L \Lambda}}\left(
    1 - \frac{\beta_1}{\beta_0^2}\frac{\log\log (\frac{1}{L \Lambda})^2}{\log (\frac{1}{L \Lambda})^2}
  \right) 
\end{align}
with $\beta_0 = (22-4\NfW)/3$ and $\beta_1=(136-64\NfW)/3$, we obtain
\begin{align}
  & {\cal M} = 8 \pi \sqrt{\frac{2 Z_{\rm bion}(g)}{g^2L^2}}
  \\
  & \sim  \frac{\phi^{-\frac{7}{8}}}{L} \Big(\log\frac{1}{L\Lambda}\Big)^{\!\!\frac{25}{8}-\frac{\beta_1}{4\beta_0}}(L\Lambda)^{\beta_{0}/2} 
  \exp\Bigg( \!\! - \Big( \scalebox{0.8}{
  $\displaystyle \sqrt{\frac{2\pi^2(\NfW-1)}{3}} + \sqrt{\frac{\NfW\phi}{\pi}}$ 
  } \Big) 
  \sqrt{\beta_0\log\frac{1}{L\Lambda}}\ \Bigg)\,, 
  \label{eq:MassGap}
\end{align}
where we have substituted $c=\sqrt{\frac{\NfW-1}{3}}$ for $N=2$ \cite{Kirkman:1981ck} and used $\Lambda$ in place of $\LQCD$ 
in order not to clutter the notation. 
While \eqref{eq:MassGap} looks fairly complicated, 
it is not difficult to see the leading behavior at $L\Lambda\ll 1$. 
For $\NfW=2$, we get $\beta_0=14/3$ and ${\cal M}\sim \frac{1}{L}(L\Lambda)^{\beta_0/2}\propto L^{4/3}\to 0$ 
as $L\to 0$.  For $\NfW=4$, we get $\beta_0=2$ and the leading powers of $L$ in \eqref{eq:MassGap} 
cancel exactly: the subleading factors then yield 
\begin{align}
  {\cal M} \sim \Big(\log\frac{1}{L\Lambda}\Big)^{\!\!\frac{65}{8}} 
  \exp\left(- \Big( 2\pi+\sqrt{\frac{8 \phi}{\pi}}\,\Big) \sqrt{\log\frac{1}{L\Lambda}}\ \right) 
  \quad \to 0\quad {\rm as}\quad L\Lambda\to 0\,.
\end{align}
The mass gap therefore vanishes in both cases in the limit $L\Lambda \to 0$. 
This conclusion is the same as that for $\phi=0$ \cite{Anber:2011de}, but 
the magnitude of ${\cal M}$ is by orders of magnitude smaller, owing to the new factor $\exp\big(- \sqrt{\frac{8\phi}{\pi}}\sqrt{\log \frac{1}{L\Lambda}} \,\big)\ll 1$. 
The string tension $\gamma$ associated with the ``spatial'' Wilson loop in $\RR^3$ is 
also diminished at $\phi\ne 0$ by the same factor, because the latter is related to the mass gap as $\gamma\sim \frac{g^2}{L} {\cal M}$ \cite{Polyakov:1976fu,Anber:2011de}. 
Such an exponential suppression of non-perturbative physics is 
the main conclusion of this subsection. 

Final remark is in order concerning the low-energy limit of the theory. 
The photons will pick up one of the $N$ degenerate minima of the bion-induced potential and the ground state will then break  
the $\ZZ_N$ shift symmetry spontaneously \cite{Unsal:2007vu,Unsal:2007jx}.%
\footnote{This $\ZZ_N$ symmetry must not be confused with the $\ZZ_N$ center symmetry, which is unbroken for $0\leq \phi \leq \phi_c$ at small $L$.}  
Since fermions and photons are both massive, there are no massless particles in the spectrum and the IR limit is trivial, as opposed to the case $\phi=0$ where 
the spectrum contains {\it massless} fermions with unbroken chiral symmetry. 

Other interesting topics not discussed in this subsection, such as neutral bions and bion-(anti-)bion molecules at $\phi\ne 0$ as well as 
their relevance to the resurgence theory and trans-series \cite{Argyres:2012vv,Argyres:2012ka}, are deferred to future work.


\subsection{Phenomenological gluonic potential}
\label{sec:NP}

\begin{figure}[t]
  \centerline{ 
  \includegraphics[width=\textwidth]{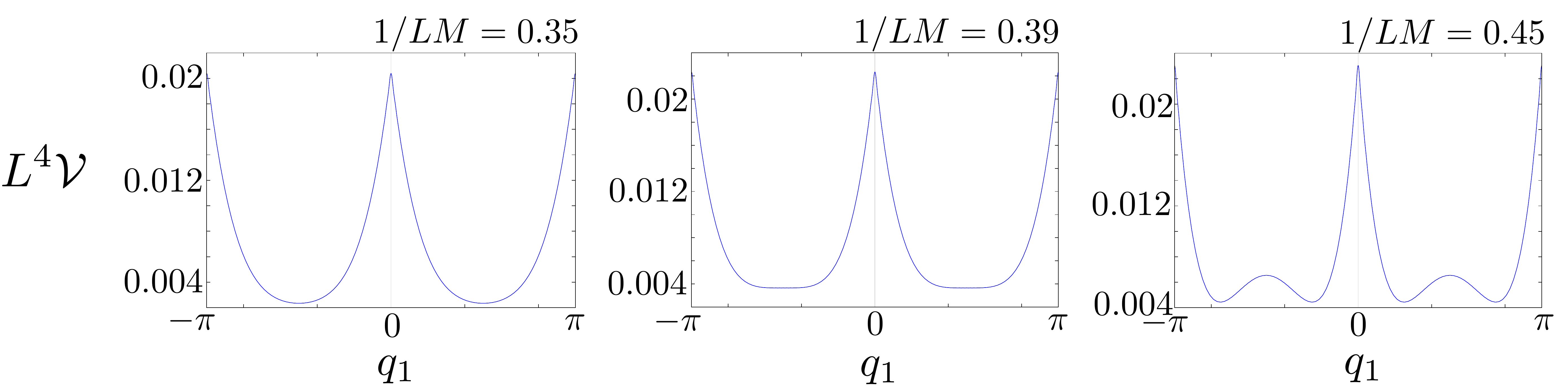}
  }
  \vspace{-.5\baselineskip}
  \caption{\label{ymP} 
  The non-perturbative potential ${\cal V}_{\YM}^{\rm np}$ for $N_{c}=2$ as a function of $q_{1}$ 
  for $1/LM = 0.35$, $0.39$, and $0.45$. A second-order phase transition occurs at $1/LM\simeq 0.390$.  
  }
\end{figure}

In Section \ref{sec:OL} we worked out the phase structure of {\qcdad} on small $\SS^1$ using perturbation theory. 
This method cannot be extended to the region $L\gtrsim \LQCD^{-1}$ where the coupling is strong. 
To gain some insight into the center symmetry realization over the whole range $0<L<\infty$, we shall in the following 
use a phenomenological gluonic potential that mimics characteristics of Yang-Mills theory as much as possible. 
Of course such a potential is not unique at all (see e.g., Ref.~\cite[Section 2]{Kashiwa:2013rmg} for comparisons). 
Here we require the potential to satisfy the following two conditions: 
\begin{itemize}
  \item It should agree with the one-loop potential ${\cal V}_{\YM}(N_{c}, L; \{q\})$ in \eqref{gc} for sufficiently small $L$.   
  \item It should reproduce the confinement/deconfinement phase transition of pure Yang-Mills theory 
  at some scale $L^{-1}=T_{\rm d} \sim \LQCD$, with the correct order of transition.
\end{itemize}
We take the following form that fulfills both requirements \cite{Meisinger:2001cq,Nishimura:2009me}: 
\begin{align}
  {\cal V}_{\YM}^{\rm np}(N_{c}, M, L ; \{q\} )
  &= - \frac{2}{\pi^{2}L^{4} } 
  \sum_{n=1}^{\infty} \frac{1}{n^4} 
  \left(1 - \frac{M^{2}L^{2}}{4} n^{2} \right) 
  \sum_{i,j=1}^{N_{c}} 
  \Big( 1 - \frac{1}{N_{c}} \delta_{ij} \Big)
  \cos( nq_{ij}) \,,
  \label{gcnp}
\end{align}
where $M\sim \LQCD$ is a non-perturbative mass scale that controls the deconfinement transition temperature. 
One can easily check that ${\cal V}_{\YM}^{\rm np}$ reduces to \eqref{gc} as $ML\to 0$.  Moreover 
it exhibits a second-order transition for $N_c=2$ and a first-order transition for $N_c=3$, in agreement with 
lattice simulations. In Figure \ref{ymP} the potential for $N_c=2$ is shown: 
the transition from a confining phase at small $L^{-1}$ to a 
deconfined phase at large $L^{-1}$ occurs at $L^{-1}\equiv T_{\rm d}^{\rm YM} \simeq 0.390 M$.  

From now, we discuss the center symmetry realization for $N_{c}=2$ and $\NfD=1$ 
in the space of $1/L$, $m$ and $\phi$ with this effective potential:   
\begin{equation}
  \label{eq:VNc2NfD1}
  {\cal V}(N_{c}=2, \NfD=1) = {\cal V}_{\YM}^{\rm np}(N_{c}=2, M, L; \{q\}) + 
  {\cal V}_{\adj}(N_{c}=2, \NfD=1, L, m, \phi; \{q\})\,, 
\end{equation}
with ${\cal V}_{\adj}$ defined in \eqref{adjc}. This extends our analysis for small $L$ 
in Section \ref{sec:OL} to the entire domain $0<L<\infty$. Our main motivation here is two-fold:  
First, we aim to grasp how the center-changing transition at small $L$ is related to the finite-temperature 
deconfinement transition at $L^{-1}=T_{\rm d}$ and $\phi=\pi$; Secondly, we explore by changing $m$ 
how the phase structure of  massless {\qcdad} $(m=0)$ evolves into that of 
the pure Yang-Mills theory $(m=\infty)$. 
Spontaneous breaking of chiral symmetry will be neglected for the moment; 
it is the main subject of Section \ref{sec:chisym}. 

\begin{figure}[t]
  \centerline{
  \includegraphics[width=0.9\textwidth]{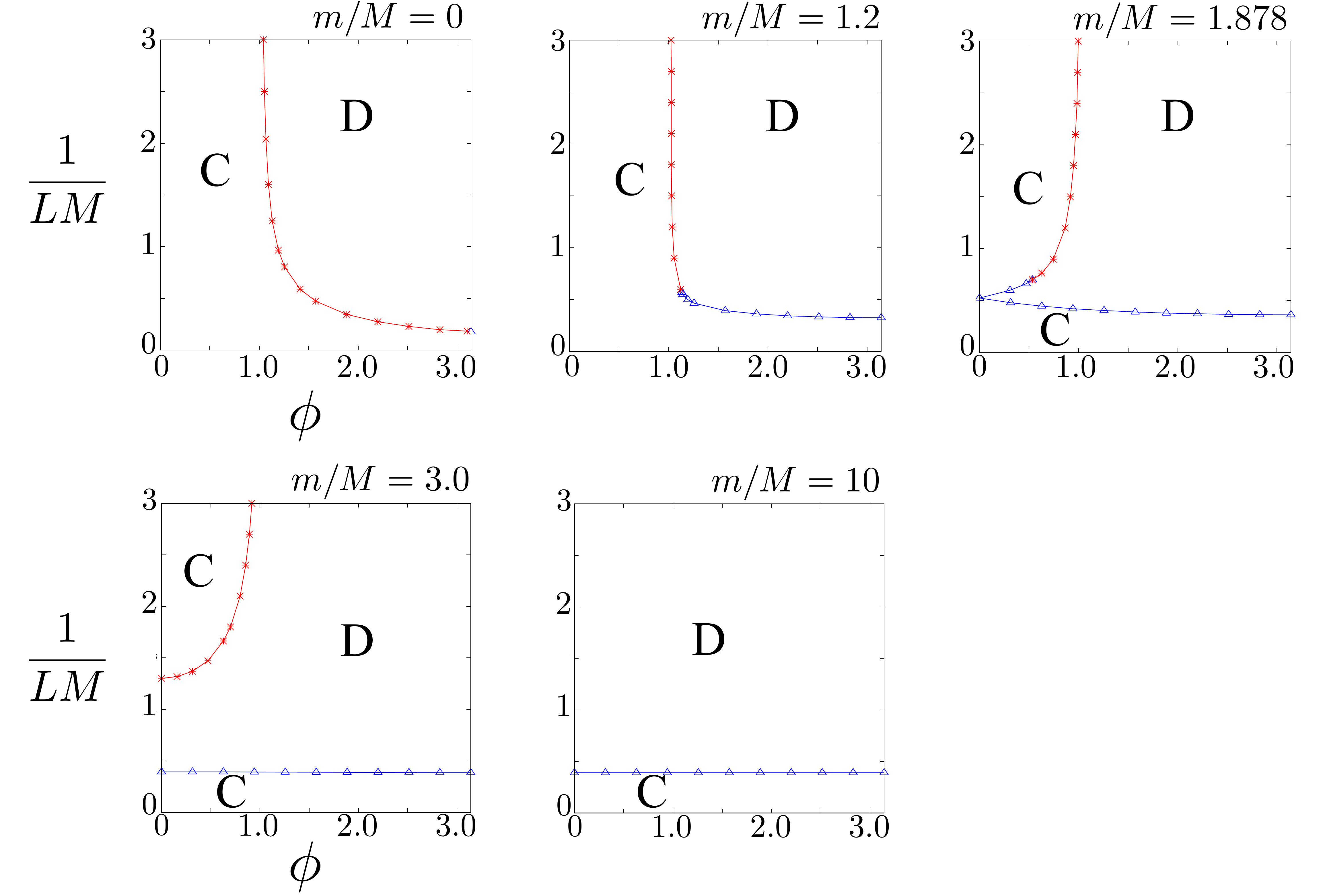}
  }
  \caption{\label{phase1} 
  Phase diagrams for $N_c=2$ and $\NfD=1$ with varying $m/M$. 
  The blue line with triangles ($\triangle$) denotes a second-order phase transition, and 
  the red line with asterisks ($\ast$) a first-order phase transition. 
  Symbols C and D are defined as before.  
  Spontaneous chiral symmetry breaking is not considered at this stage. 
  }
\end{figure}

In Figure~\ref{phase1} we present the numerically obtained phase diagrams on $1/L$--$\phi$ space with varying $m$.  
At $m=0$ (top left in Figure~\ref{phase1}), the finite-temperature $(\phi=\pi)$ phase transition is found to be second order and occurs at 
$L^{-1} =T_{\rm d}^{\NfD=1} \simeq 0.177M$.  In comparison to $T_{\rm d}^{\rm YM} \simeq 0.390 M$, the transition 
temperature has been reduced by a factor of $0.45$ by inclusion of massless fermions. 
It is notable that the deconfinement phase transition seems to become \emph{first order} as soon as $\phi$ is detuned from $\pi$. 
At $m=0$ the confining phase at low temperature $\frac{1}{L}\ll M$ is continuously connected 
to the gauge-symmetry-broken phase at $\frac{1}{L}\gg M$. This is not a contradiction, because 
there is no gauge-invariant order parameter that characterizes the Higgs phenomenon. For 
$\frac{1}{L}\gg M$, the previous one-loop analysis is valid (recall \eqref{eq:table_2_1}) and 
we expect the center-changing transition to occur at $\phi =0.326\pi\simeq 1.024$ --- 
indeed this is what we see in the figure. 

As $m$ is increased from zero, the second-order transition line emanating from the $\phi=\pi$ axis 
extends into the interior of the phase diagram. Passing through a tricritical point, it becomes first order.%
\footnote{A similar tricritical point has been found numerically in deformed Yang-Mills theory \cite{Ogilvie:2012fe}. }  
At $m/M\simeq 1.878$, the second-order line hits the $\phi=0$ axis, and for larger $m$ the confined 
phase is separated into two domains.  As $m$ is further increased, the confined (gauge-symmetry-broken) phase 
is pushed to higher and higher $\frac{1}{LM}$ and disappears from the figure; 
the phase diagram finally reduces to that of pure Yang-Mills theory, which is independent of $\phi$.  
Such a change of the phase structure is consistent with the expected decoupling of fermions for $\frac{1}{L}\lesssim m$.  
If $m$ is interpreted as mimicking the \emph{constituent} quark mass, then an analog of the above 
behavior may well arise in the phase diagram of full-fledged $\qcdad$ as well.  In particular, 
the fact that heavy fermions cannot sustain the center-symmetric phase at intermediate $L$ could be 
detrimental to the concept of ``adiabatic continuity'' from small $L$ 
to large $L$ \cite{Argyres:2012vv,Argyres:2012ka,Dunne:2012ae,Dunne:2012zk} and to the large-$N$ 
volume independence, so we will revisit this issue in Section \ref{sec:chisym}. 

\begin{figure}[t]
  \centerline{
  \includegraphics[width=0.95\textwidth]{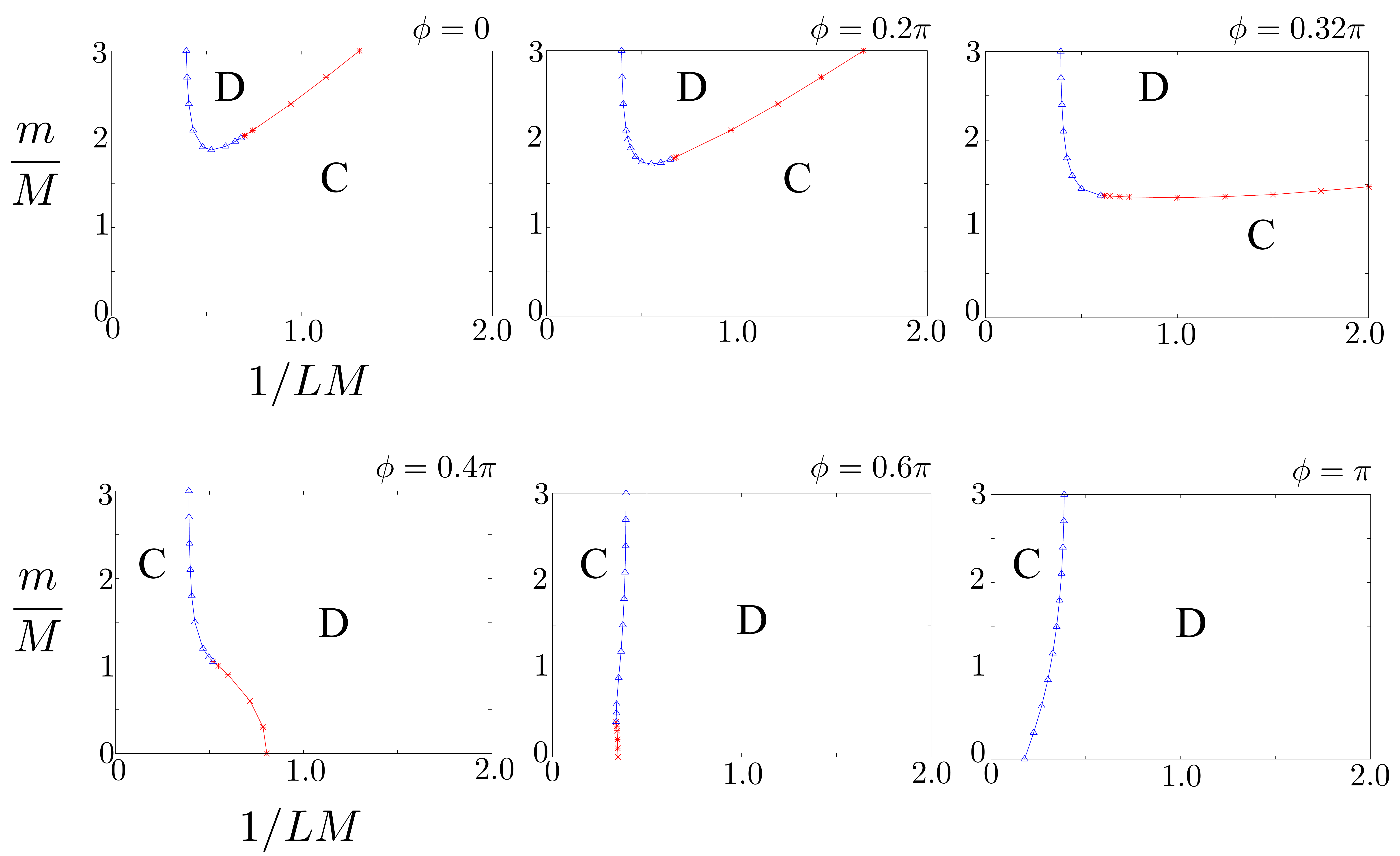}
  }
  \caption{\label{Phi_v}
  Phase diagram for for $N_c=2$ and $\NfD=1$ with 
  $\phi \in \{0,~0.2\pi,~0.32\pi,~0.4\pi,~0.6\pi,~\pi\}$.  
  The blue line with triangles ($\triangle$) denotes a second-order phase transition, and 
  the red line with asterisks ($\ast$) a first-order phase transition. 
  }
\end{figure}

Finally, in Figure~\ref{Phi_v}, we depict the phase diagram of the same theory, but this time on the 
$(\frac{1}{LM}\,, \frac{m}{M})$ plane for a variety of $\phi$. 
These figures clearly show how the phase structure changes with 
the boundary conditions.  In the top left panel of Figure~\ref{Phi_v}, the transition curve reaches its lowest point with 
$m/M=1.878$, in agreement with the top right panel of Figure~\ref{phase1}.     
Keeping track of the evolution of the phase diagram in Figure~\ref{Phi_v}, one sees a qualitative 
global change of the phase diagram at $\phi\simeq 0.326\pi$, 
which is exactly the phase transition point in one-loop perturbation theory (cf.~\eqref{eq:table_2_1}). For $\phi$ greater than this value, 
even massless fermions cannot maintain a center-symmetric phase at large $\frac{1}{LM}$.  
Direct comparison of these model predictions and lattice simulations for $\phi=0$ \cite{Cossu:2009sq,Cossu:2013ora} 
will be postponed to Section \ref{sec:chisym} where dynamical chiral symmetry breaking is taken into account.  

A caveat is in order concerning the validity of the mean-field approximation we used.%
\footnote{This caveat also pertains to Section \ref{sec:chisym} where chiral and center symmetry breaking are simultaneously considered.}  
In effective models for QCD and pure Yang-Mills theory, it has been known that 
treating the Polyakov-loop matrix as a naive mean field is quite subtle and can even cause 
pathological behaviors \cite{Abuki:2009dt}. For instance, the relation $\langle\Tr_{\adj}\Omega\rangle = |\langle\Tr \Omega\rangle|^2-1$,  
valid in a naive mean-field approximation, is at variance with the lattice data where vanishingly small values are reported 
for both $\langle\Tr \Omega\rangle$ and $\langle\Tr_{\adj} \Omega\rangle$ at $T<T_{\rm d}$ \cite{Gupta:2007ax}.   
Presumably one can expect this approximation to work accurately only under special circumstances, e.g., in the large-$N$ limit 
and in the weak-coupling regime at small $\SS^1$. 
Attempts to fix problems, e.g., through the use of Weiss mean-field approximation, are reported 
in the literature \cite{Abuki:2009dt,Zhang:2010kn,Sasaki:2012bi,Ruggieri:2012ny}. It seems to be 
a very promising future direction to extend our work using such more refined approximation schemes.

\section{Impact of chiral symmetry breaking}
\label{sec:chisym}

In this section we will incorporate the effect of dynamical chiral symmetry breaking to  
study the interplay of center and chiral symmetry on the phase diagram of $\qcdad$.  
In Section \ref{sec:prelim} we review global symmetries of {\qcdad} and comment on 
restrictions on the possible patterns of symmetry breaking. The insights obtained here 
are used for an appropriate model building in Section \ref{sec:PNJL}, where 
we analyze an NJL-type model with twisted boundary conditions using a 
high-temperature expansion. In Section \ref{sec:numerics} the PNJL model is solved numerically 
and its phase diagram is presented.

\subsection[Chiral symmetry in $\qcdad$]
{\boldmath Chiral symmetry in $\qcdad$}
\label{sec:prelim}

In $\qcdad$ on $\RR^4$ in the chiral limit, the flavor symmetry of adjoint fermions is given by 
$\U(1)_{\rm A}\times \SU(\NfW)$ with $\NfW=2\NfD$, which is larger than the standard symmetry 
$\U(1)_{\rm A}\times \U(1)_{\rm B}\times \SU(\NfD)_{\rm R}\times\SU(\NfD)_{\rm L}$ 
owing to the reality of the adjoint representation \cite{Leutwyler:1992yt,Kogut:2000ek}.%
\footnote{We only consider even $\NfW$ (cf.~Section \ref{sec:OL}). }  
Instantons further break $\U(1)_{\rm A}$ down to $\ZZ_{2 N_c \NfW}$ explicitly, for the case of 
$\SU(N_c)$ gauge theory \cite{Creutz:2007yr}. The spontaneous breaking of continuous 
and discrete chiral symmetries can be probed by $\langle \Tr \lambda^f \lambda^g \rangle$ 
and $\langle \underset{f,g}{\det} \Tr \lambda^f \lambda^g \rangle$, respectively, 
where $\lambda^f~(f=1,\dots,\NfW)$ are adjoint Weyl fermions and `$\Tr$' denotes a trace 
over color indices. On small $\SS^1$ with PBC, $\ZZ_{2 N_c \NfW}$ is spontaneously 
broken to $\ZZ_{2\NfW}$ yielding $N_c$ degenerate vacua \cite{Unsal:2007jx,Unsal:2008eg}.   

Now we consider $\qcdad$ on $\RR^3\times \SS^1$ with twisted boundary conditions ($0\leq \phi\leq \pi$).  Since 
imposing a twist $\phi$ on Dirac fermions $\Psi$ as in \eqref{eq:fbc} is equivalent to imposing a twist $\phi$ 
on half of the Weyl fermions and $-\phi$ on the other half, it follows that the flavor symmetry is {\it explicitly broken} by $\phi$ down to 
$\U(1)_{\rm A}\times \U(1)_{\rm B}\times \SU(\NfW/2)_{\rm R}\times\SU(\NfW/2)_{\rm L}$, 
except for $\phi/\pi \in \ZZ$ at which the full symmetry $\U(1)_{\rm A}\times \SU(\NfW)$ is recovered. 
This can also be seen as follows: a twist is equivalent to a constant background $\U(1)$ gauge field $A_4=\phi/L$, 
which spoils the reality of the adjoint Dirac operator and reduces the flavor symmetry down to that of fermions 
in a complex representation. Since the enlarged symmetry is broken, 
it is more appropriate to use $\langle \Tr \bar\Psi\Psi \rangle$ instead of 
$\langle \Tr \lambda \lambda \rangle$ as an order parameter for flavor symmetry breaking. 

A rough picture on the chiral phase structure is as follows. 
For $L\gg \LQCD^{-1}$, continuous chiral symmetry is spontaneously broken 
(for $\NfW$ outside the conformal window) for all $0\leq\phi\leq \pi$. 
For $L\ll\LQCD^{-1}$, by contrast, the running coupling is small and chiral symmetry will be restored.  
Thus there must be a chiral transition at some $L=L_{\chi}(\phi)$ for any $0\leq \phi\leq \pi$. 

As asymptotic freedom requires $\NfD<11/4$ (see \eqref{eq:betafunc}), either $\NfD=1$ or $2$ is possible. 
Let us discuss the two cases separately. 
\begin{itemize}
  \item 
  $\NfD=1$ \,:  Non-anomalous global symmetry is 
  $\displaystyle \frac{(\ZZ_{4N_c})_{\rm A}\times\SU(2)}{\ZZ_2}$ for $\phi=0$ or $\pi$, and 
  $\displaystyle \frac{(\ZZ_{4N_c})_{\rm A}\times\U(1)_{\rm B}}{\ZZ_2}$ for $0<\phi<\pi$.%
  \footnote{The common subgroup $\ZZ_2$ is factored out to avoid double counting of symmetries. 
  Same for $\NfD=2$.} 
  The thermal chiral transition at $\phi=\pi$, driven by $\langle\Tr\bar\Psi\Psi\rangle\ne 0$, should be 
  either first order, or second order in the $\O(3)$ universality class \cite{Basile:2004wa,Basile:2005hw}. 
  \item 
  $\NfD=2$ \,:  Non-anomalous global symmetry is 
  $\displaystyle \frac{(\ZZ_{8N_c})_{\rm A}\times\SU(4)}{\ZZ_4}$ for $\phi=0$ or $\pi$, and 
  $\displaystyle 
  \frac{(\ZZ_{8N_c})_{\rm A}\times \U(1)_{\rm B}\times \SU(2)_{\rm R}\times\SU(2)_{\rm L}}{\ZZ_4 \times (\ZZ_2)_{{\rm R}+{\rm L}}}$ 
  for $0<\phi<\pi$. 
  The thermal chiral transition at $\phi=\pi$, driven by $\langle\Tr\bar\Psi\Psi\rangle\ne 0$, is likely to be second order for $N_c=3$ 
  according to lattice data \cite{Karsch:1998qj,Engels:2005te}, which should belong to the $\SU(4)/\SO(4)$ 
  universality class \cite{Basile:2004wa,Basile:2005hw}. 
\end{itemize}

As one can see above, $\NfD=1$ {\qcdad} is distinct from $\NfD=2$ in  
that the chiral condensate $\langle\Tr\bar\Psi\Psi\rangle$ 
breaks no continuous symmetry at $\phi\ne 0$. We now argue that 
the pattern of \emph{discrete} chiral symmetry breaking 
for $\NfD=1$ is dependent on the background holonomy   
through the existence of yet another condensate $\langle\det \Tr \lambda^f \lambda^g \rangle$. 
The latter assumes a nonzero VEV if either of the following two conditions is met:
\begin{enumerate}
  \item[(1)] If $\langle\Tr\bar\Psi\Psi\rangle$ is nonzero, it naturally induces 
  a nonzero VEV of $\langle\det \Tr \lambda^f \lambda^g\rangle$ because 
  there is no symmetry that protects the latter from taking a nonzero value. 
  \item[(2)] If the holonomy is center-symmetric, then 
  instantons split into $N_c$ constituent monopoles. The fermionic zero modes 
  attached to each monopole contribute to making $\langle\det \Tr \lambda^f \lambda^g\rangle$ nonzero \cite{Davies:1999uw,Davies:2000nw}.%
  \footnote{Although this picture is justified only in a small-$L$ semiclassical domain,  
  we suspect that the determinantal condensate is generally supported by a 
  nontrivial holonomy. Indeed, a recent preprint \cite{Bergner:2014saa} reports that 
  $\langle\lambda\lambda\rangle$ in SYM goes to zero at approximately 
  the same temperature as the deconfinement transition, which corroborates our 
  point of view.} 
\end{enumerate}

The above consideration suggests that, at $\phi\ne0$, 
we should take into account the background Polyakov loop 
to figure out the correct universality class of the chiral transition 
driven by $\langle\Tr\bar\Psi\Psi\rangle$.
\begin{itemize}
  \item In a center-symmetric phase with $\langle\PF\rangle=0$, 
  the axial symmetry $(\ZZ_{4N_c})_{\rm A}$ is spontaneously broken to $\ZZ_{4}$ 
  through $\langle\det \Tr \lambda^f \lambda^g\rangle\ne 0$. Therefore 
  $\langle\Tr\bar\Psi\Psi\rangle$ serves as an order parameter of the 
  symmetry breaking $\ZZ_{4}\to \ZZ_{2}$\,. The chiral transition, if continuous, 
  will belong to the 3d Ising universality class. 
  \item In a deconfined phase with $\langle\PF\rangle\ne 0$, 
  $\langle\det \Tr \lambda^f \lambda^g\rangle$ will go to zero 
  simultaneously with $\langle\Tr\bar\Psi\Psi\rangle$ at the chiral transition. 
  Therefore the pattern is $\ZZ_{4N_c}\to\ZZ_{2}$\,.  
  The chiral transition, if continuous, 
  will belong to the same universality class as spin systems in 3d 
  with $\ZZ_{2N_c}$ symmetry breaking. 
\end{itemize}
This issue will be taken up again in Section \ref{sec:numerics} 
when we discuss the phase diagram of {\qcdad}. 

A comment is in order concerning the $\U(1)_{\rm B}$ baryon number symmetry. 
As is well known, {\qcdad} enters a superfluid phase 
with diquark condensate $\langle\Tr\Psi\Psi\rangle\ne0$ 
for high \emph{real} chemical potential \cite{Kogut:2000ek,Kanazawa:2011tt}. However, 
it does not occur for \emph{imaginary} chemical potential (i.e., $\phi\ne 0$) because it is forbidden  
by the Vafa-Witten theorem \cite{Vafa:1983tf} for any nonzero bare mass $m$. 
The argument goes as follows. If $\U(1)_{\rm B}$ were spontaneously broken, there will be a massless NG mode 
$M\sim \Psi\Psi$, whose correlator $\langle M(x)M^\dagger(y)\rangle$ decays more slowly than exponentially at large $|x-y|$. 
On the other hand, (i) if the path-integral measure is positive definite, and (ii) if the Dirac operator is anti-hermitian,  
one can prove that the fermion propagator $S(x,y)$ in an arbitrary gauge field background 
decays at least as fast as $\ee^{-m|x-y|}$ at large distances \cite{Vafa:1983tf}. 
Thus $|\langle M(x)M^\dagger(y)\rangle| \propto |S(x,y)|^2 \leq \ee^{-2m|x-y|}$, which contradicts the existence of a 
massless NG mode in this channel.  
As the conditions (i) and (ii) are both satisfied for {\qcdad} at any $\phi$, we conclude that 
$\U(1)_{\rm B}$ is not spontaneously broken, and $\langle\Tr\Psi\Psi\rangle=0$. 
This argument equally applies to both $\NfD=1$ and $2$. 
For $\NfD=2$, $\langle\Tr\Psi\Psi\rangle=0$ can also be proven 
by means of QCD inequalities \cite{Weingarten:1983uj,Witten:1983ut,Nussinov:1983hb}. 

Thus the diquark condensate vanishes in {\qcdad} for any $\phi$, as long as $m\ne 0$. If we define 
the chiral limit as $m\to +0$, this conclusion should remain valid in the chiral limit as well.%
\footnote{Note that there are other ways to approach the chiral limit; for example, in the presence of a diquark source $j\Psi\Psi$, 
the ground state may depend on which of the limits $m\to 0$ and $j\to 0$ were taken first \cite{Kogut:2000ek,Kanazawa:2011tt}. 
The arguments presented above do not apply to the case with $j\ne 0$.} 
This allows us to take $\langle\Tr\bar\Psi\Psi\rangle$ 
as the primary order parameter of flavor symmetry breaking 
in {\qcdad} within the analysis of the phase diagram.

\subsection{PNJL model: an analytical study}
\label{sec:PNJL}

In this subsection we study the $\phi$-dependence of dynamical chiral symmetry breaking analytically. 
For simplicity, the bare mass of fermions is set to zero.  

From here on, we shall concentrate on the $\NfD=1$ case, for the sake of simplicity and to evade the conformal window.  
We want to investigate chiral symmetry realization for general $\phi$ using an effective model. For this purpose, 
it is essential that the model reflects the special symmetries of {\qcdad} correctly.%
\footnote{This point was not stressed in earlier works \cite{Nishimura:2009me,Kashiwa:2013rmg}; 
see Appendix \ref{app:modeltable} for a detailed comparison of chiral models in the literature.} 
The form of four-fermion interaction that respects 
$\SU(2\NfD)$ symmetry of {\qcdad} has been determined in Ref.~\cite{Zhang:2010kn}. For $\NfD=1$, 
two kinds of interaction have been identified: 
\begin{subequations}
  \begin{align}
    {\cal L}_{\U(2)} & = \frac{G}{2}\big[ (\bar\Psi \Psi)^2 + (\bar\Psi i\gamma_5 \Psi)^2 + |\overline{\Psi^C}\gamma_5\Psi|^2 + |\overline{\Psi^C}\Psi|^2 \big] 
    \label{eq:Lu2}
    \\
    {\cal L}_{\SU(2)} & = \frac{G}{2}\big[ (\bar\Psi \Psi)^2 - (\bar\Psi i\gamma_5 \Psi)^2 + |\overline{\Psi^C}\gamma_5\Psi|^2 - |\overline{\Psi^C}\Psi|^2 \big] 
    \label{eq:Lsu2}
  \end{align}
\end{subequations}
where $G$ is a coupling constant of dimension $[\text{mass}]^{-2}$,  
$\Psi$ is the adjoint Dirac spinor, $\Psi^C\equiv C\bar{\Psi}^T$ is the charge-conjugated spinor, 
and the color indices are assumed to be properly contracted. 
Both \eqref{eq:Lu2} and \eqref{eq:Lsu2} entertain $\SU(2)$ symmetry and, for $0<\phi<\pi$, break it down to $\U(1)_{\rm B}$ explicitly, 
in agreement with the expectation for $\qcdad$. 
Their difference is that, as the notation suggests, ${\cal L}_{\U(2)}$ is invariant under $\U(1)_{\rm A}$ whereas ${\cal L}_{\SU(2)}$ is not; the latter breaks 
$\U(1)_{\rm A}$ down to $\ZZ_4$.  On the other hand, 
the correct unbroken axial symmetry for $\NfD=1$ is $\ZZ_{4N_c}$, as remarked above. This means neither ${\cal L}_{\U(2)}$ 
nor ${\cal L}_{\SU(2)}$ has the correct axial symmetry. Is it important for us? Probably yes, because the 
axial anomaly can change the order of the chiral phase transition, and in case of continuous transition, can change 
the corresponding universality class, as revealed in the renormalization group analysis \cite{Pisarski:1983ms,Basile:2004wa,Basile:2005hw}.  
However, such a detailed description of the chiral transition is beyond the scope of this paper. Instead, in what follows, 
we will just be satisfied with demonstrating the chiral phase structure on $L^{-1}$--$\phi$ plane. 
For practical calculations we would like to use a linear combination of ${\cal L}_{\U(2)}$ and ${\cal L}_{\SU(2)}$ 
in order to mimic the effect of anomaly. Namely, we consider a chiral effective model whose fermionic part is given by
\begin{align}
  {\cal L}_{\NfD=1} & = \Tr [ \bar\Psi D_{\adj}\Psi ] + 
  \frac{1}{2}\big({\cal L}_{\U(2)} + {\cal L}_{\SU(2)} \big) 
  \\
  & = \Tr [ \bar\Psi D_{\adj}\Psi ] + 
  \frac{G}{2}\big[ (\bar\Psi \Psi)^2 + |\overline{\Psi^C}\gamma_5\Psi|^2 \big]
  \,, 
  \label{eq:LNJL}
\end{align}
where $D_{\adj}\Psi \equiv \gamma_\mu \big(\der_\mu \Psi + i [A_\mu, \Psi]\big)$ and 
$A_\mu=\delta_{\mu 4}A_4$ with $A_4$ given in \eqref{eq:A4parametrization}.  
The bare mass is set to zero. This is a variant of the so-called PNJL model \cite{Fukushima:2003fw}. 
In the following, we treat the gauge sector along the lines of Section \ref{sec:NP} using 
the phenomenological gluonic potential ${\cal V}_{\YM}^{\rm np}$. If the coupling $G$ is zero, 
we are brought back to the model of Section \ref{sec:NP}, whose total effective potential  
is already given in \eqref{eq:VNc2NfD1} for the case of $N_c=2$. 
In Appendix \ref{app:modeltable}, the present model is juxtaposed with other 
chiral effective models in the literature for completeness.  

Now, we switch on $G>0$ and investigate its impact on the phase diagram. 
The model \eqref{eq:LNJL} can be solved in the mean-field approximation 
following a standard procedure \cite{Klevansky:1992qe,Hatsuda:1994pi}. 
From Section \ref{sec:prelim}, we expect 
$\langle\Tr \overline{\Psi^C}\gamma_5\Psi \rangle=0$ in {\qcdad}, and will assume this 
in the current effective model as well. Then the total thermodynamic potential 
at the mean-field level reads%
\footnote{At the mean-field level, the thermodynamic potential does not depend on whether we have 
included the diquark channel in the four-fermi interaction or not. This is simply because we have no 
diquark condensate in the present situation. However, once we go beyond the mean-field approximation 
and incorporate mesonic fluctuations, it matters to have the four-fermion interaction \eqref{eq:LNJL} 
with the correct symmetry of {\qcdad}.} 
\begin{align}
  & {\cal V}_{\rm tot}(N_c,\NfD=1, \phi, L, \md ; \{q\}) 
  \notag 
  \\
  & = 
  {\cal V}_{\YM}^{\rm np}(N_{c}, M, L; \{q\}) + 
  {\cal V}_{\adj}(N_{c}, \NfD=1, L, \md, \phi; \{q\}) 
  + {\cal V}_{\chi}(N_{c}, G, \Luv, \md)\,,
  \label{eq:Vtotal}
\end{align}
where $\md \equiv -G \langle\Tr \bar\Psi \Psi\rangle$ is the dynamical mass of fermions, 
${\cal V}_{\YM}^{\rm np}$ and ${\cal V}_{\adj}$ are defined in \eqref{gcnp} and \eqref{adjc}, 
respectively, and   
\begin{align}
  {\cal V}_{\chi}(N_{c}, G, \Luv, \md) 
  & = \frac{\md^2}{2G} - 2 (N_c^2-1) \!\!\!\! \int\limits_{|p|<\Luv} \!\!\!\!\! \frac{d^3p}{(2\pi)^3}\sqrt{p^2+\md^2}
  \\
  & = \frac{\md^2}{2G} - \frac{N_c^2-1}{8\pi^2} 
  \Bigg\{ \Luv (\md^2+2\Luv^2) \sqrt{\md^2+\Luv^2} 
  \notag
  \\
  & \qquad + \md^4 \log \frac{\md}{\Luv+\sqrt{\md^2 + \Luv^2}} \Bigg\}
  \,,
\end{align}
where $\Luv$ is a momentum cutoff.  

Finding the minimum of ${\cal V}_{\rm tot}(N_c,\NfD=1, \phi, L, \md ; \{q\})$ for all $0\leq \phi\leq \pi$ requires tedious numerics,  
which will be done in Section \ref{sec:numerics} for the simplest case of $N_c=2$. Before that, let us try to capture 
qualitative features of the chiral transition for varying $\phi$ by analytical means, which proves helpful in developing an intuitive 
picture for the interplay of confinement and chiral symmetry breaking. To facilitate practical calculations, 
we ``freeze'' the dynamics of the gauge sector by fixing a particular holonomy and then 
solve the chiral dynamics in this background.%
\footnote{This can be implemented in actual lattice simulations 
with the aid of double-trace deformations \cite{Unsal:2008ch}.} 
Section \ref{sec:trivialholo} deals with the case of a trivial holonomy 
$\Omega = \1$ where the model reduces to the classical NJL model but with a twisted boundary condition. 
In Section \ref{sec:confinedholo} we turn to the case of a center-symmetric holonomy.

\subsubsection{Trivial holonomy}
\label{sec:trivialholo}

Assuming a continuous chiral phase transition, $\md$ is small in the vicinity of chiral restoration, 
and the effective potential can be expanded in terms of $\md$. With inspection we get 
\begin{align}
  {\cal V}_{\chi}(N_{c}, G, \Luv, \md) & = {\rm const.} + \left(\frac{1}{2G}-\frac{N_c^2-1}{4\pi^2}\Luv^2\right)\md^2 + O(\md^4)\,.
  \label{eq:Vchiexp}
\end{align} 
In the decompactification limit $L\to\infty$, ${\cal V}_{\adj}$ in \eqref{eq:Vtotal} drops off and ${\cal V}_{\chi}$ solely determines 
the chiral symmetry realization. Since we are interested in theories in which chiral symmetry is broken on $\RR^4$, 
the coefficient of the second-order term should be negative, i.e.,%
\footnote{Chiral symmetry in this model would be unbroken on $\RR^4$ if \eqref{eq:chsb} is not satisfied. Although such a chirally-symmetric phase 
is reminiscent of the IR conformal phase of QCD with many flavors, care must be taken, because the conformality is broken by the dimensionful 
parameter $G$ \emph{explicitly}, and also because the pointlike four-fermion interaction of the NJL model is presumably  
a poor approximation to the long-range interaction in the conformal phase of QCD mediated by massless gluons. } 
\begin{align}
  G > \frac{2\pi^2}{N_c^2-1}\frac{1}{\Luv^2}\,. 
  \label{eq:chsb}
\end{align}
At finite $L$, ${\cal V}_{\adj}$ gives a nonzero contribution. 
Putting $\Omega=\1$ in \eqref{preadjc}, one finds
\begin{align}
  {\cal V}_{\adj}(N_{c}, \NfD=1, L, \md, \phi; \{q\}) 
  = (N_c^2-1) \frac{\md^2}{\pi^{2} L^{2}} \sum_{n=1}^{\infty} 
  \left( \frac{K_2(nL\md)}{n^2} \ee^{in\phi} + \text{c.c.} \right)\,.
\end{align}
A naive expansion of $K_2$ in powers of $\md$ fails, because the sum over $n$ becomes termwise divergent. 
The correct expansion in $\md$ for generic $\phi$ has been derived quite recently by Klajn \cite{Klajn:2013gxa}. 
Using formulas from Ref.~\cite{Klajn:2013gxa}, we obtain 
\begin{align}
  {\cal V}_{\adj}(N_{c}, \NfD=1, L, \md, \phi; \{q\}) 
  = & - (N_c^2-1) \left[
    \frac{4\pi^2}{3L^4}B_4 \left(\kakko{\frac{\phi}{2\pi}}\right) + \frac{1}{L^2}B_2\left(\kakko{\frac{\phi}{2\pi}}\right)\md^2 
  \right] 
  \notag 
  \\
  & + \big\{\text{higher orders in } \md \big\}\,, \hspace{-30pt}
  \label{eq:Vadexp}
\end{align}
where $B_2(x)$ and $B_4(x)$ are the Bernoulli polynomials and $\kakko{x} \equiv x - \lfloor x \rfloor \in[0,1)$ is the fractional part of $x$.%
\footnote{The leading higher-order term in \eqref{eq:Vadexp} is $O\big(\md^4\log \md^2 \big)$ for $\phi\ne 0$ 
and $O\big(|{\md}|^3 \big)$ for $\phi=0$ \cite{Klajn:2013gxa}. } 

Combining \eqref{eq:Vchiexp} and \eqref{eq:Vadexp}, we arrive at the expansion of the total potential:
\begin{multline}
  \qquad \qquad 
  {\cal V}_{\rm tot}(N_c,\NfD=1, \phi, L, \md ; \{q\})\Big|_{O\big(\md^2 \big)}
  \\
  = \left[ \frac{1}{2G}-\frac{N_c^2-1}{4\pi^2}\Luv^2
  - \frac{N_c^2-1}{L^2} B_2\left(\kakko{\frac{\phi}{2\pi}}\right) 
  \right] \md^2 \,. \qquad 
\end{multline}
We mention that this expression has been independently derived 
by Beni\'{c} \cite{Benic:2013zaa}. 

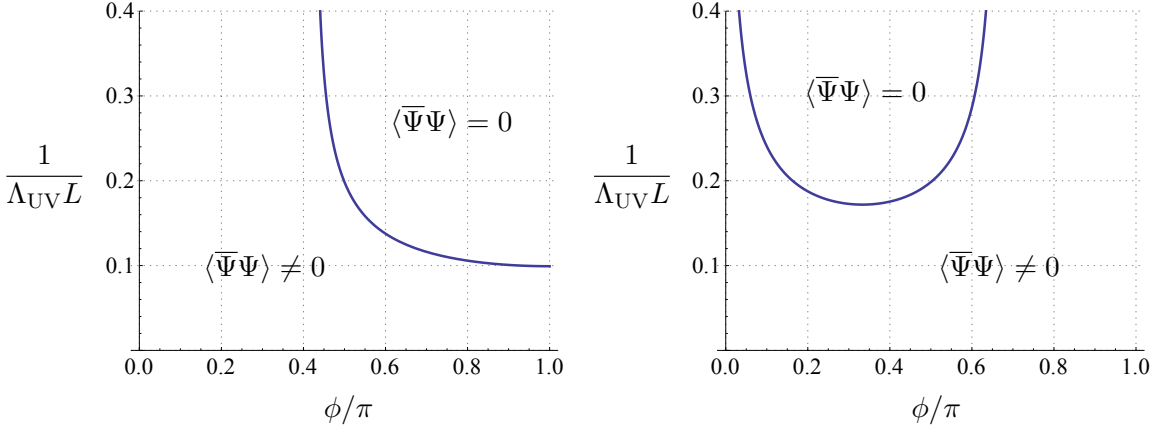
\begin{figure}[t]
  \centerline{
    \qquad \quad 
    \includegraphics[width=0.4\textwidth]{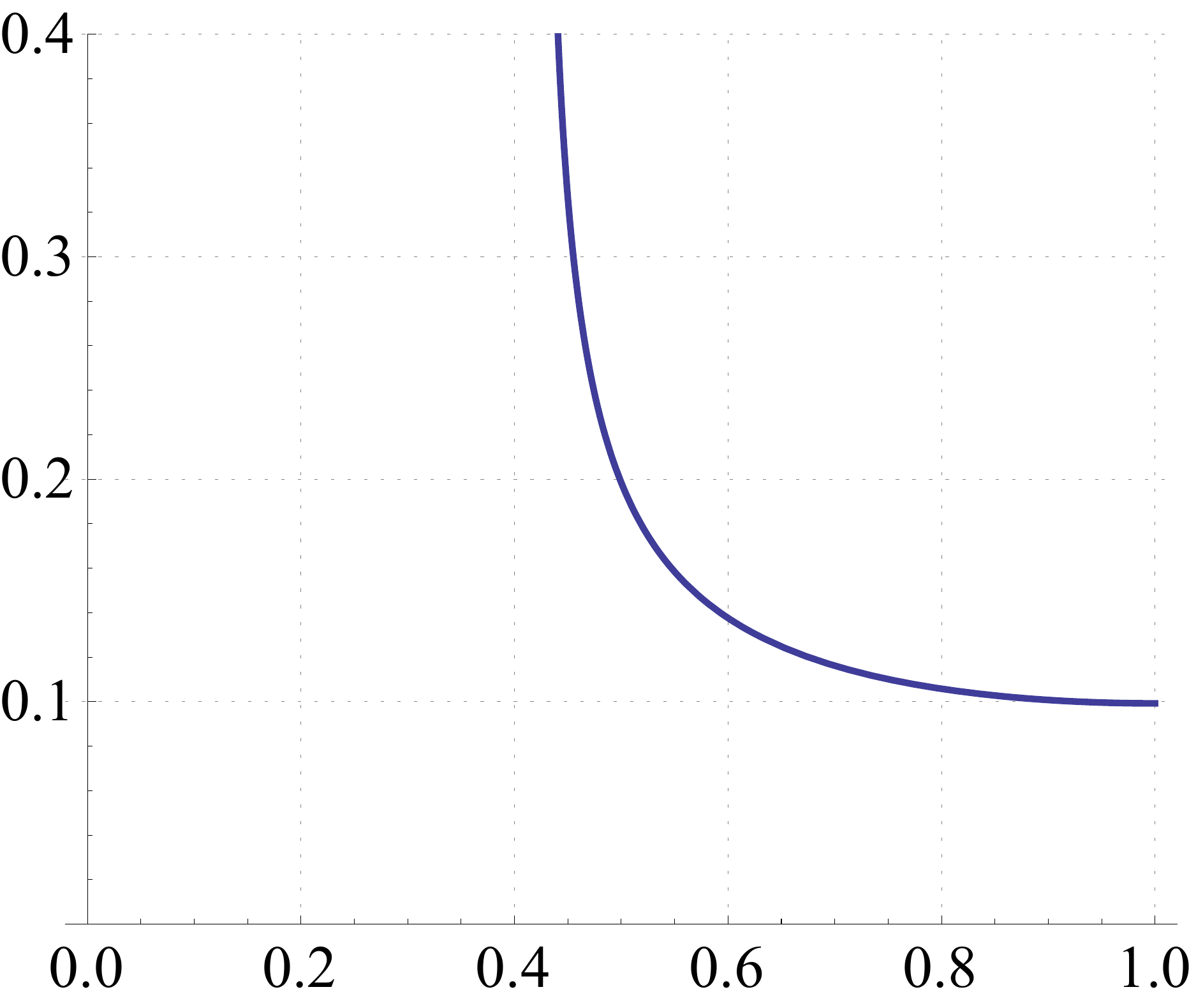}
    \qquad \qquad 
    \includegraphics[width=0.4\textwidth]{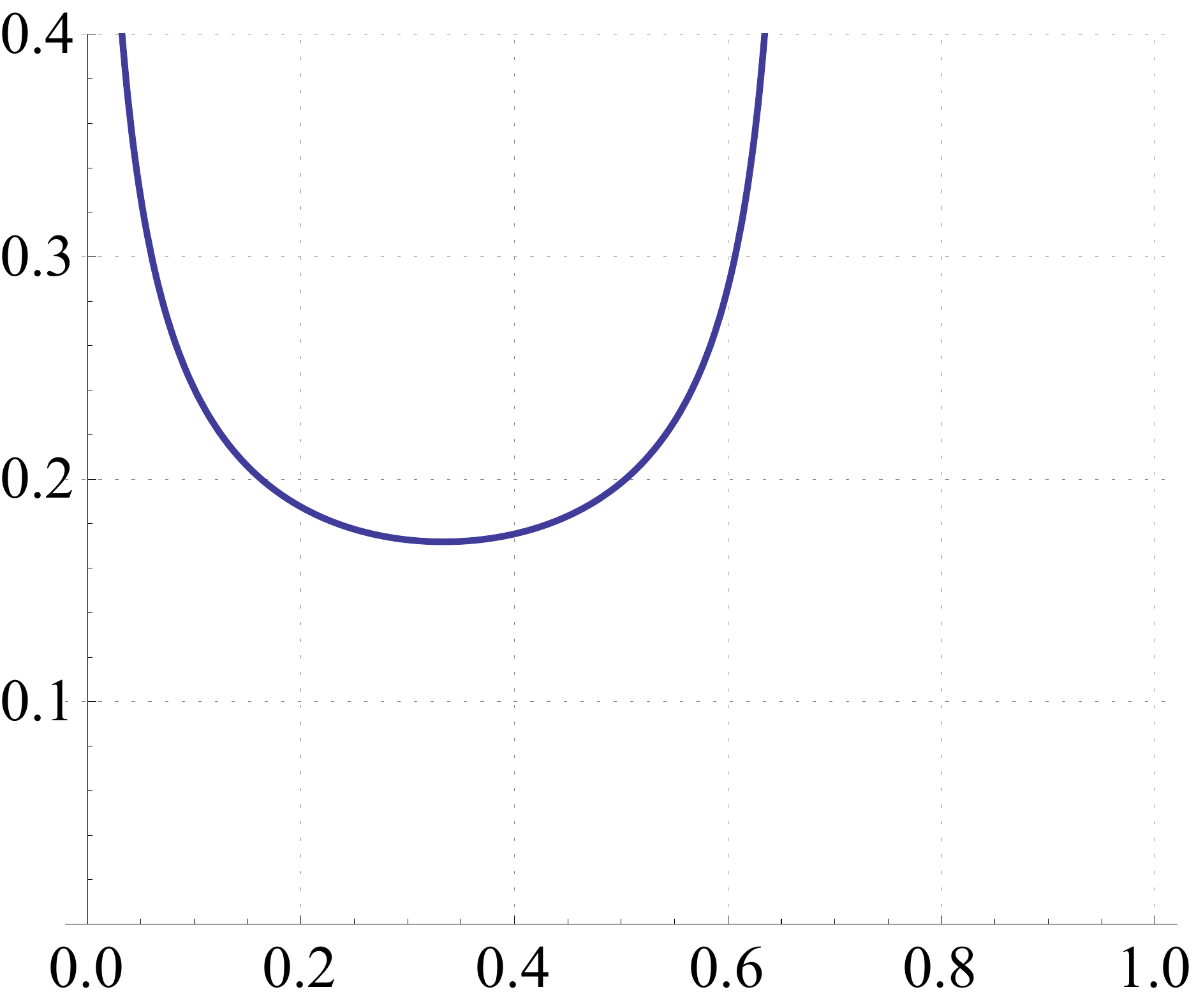}
    \put(-90,-13){$\phi/\pi$}
    \put(-310,-13){$\phi/\pi$}
    \put(-210,75){$\displaystyle \frac{1}{\Luv L}$}
    \put(-430,75){$\displaystyle \frac{1}{\Luv L}$}
    \put(-355,40){$\langle\overline{\Psi}\Psi\rangle\ne 0$}
    \put(-285,95){$\langle\overline{\Psi}\Psi\rangle= 0$}
    \put(-80,40){$\langle\overline{\Psi}\Psi\rangle\ne 0$}
    \put(-130,106){$\langle\overline{\Psi}\Psi\rangle= 0$}
  }
  \caption{\label{fg:njlphases}
    Phase structures of the NJL model for trivial holonomy {\bf (left)} and for center-symmetric holonomy {\bf (right)} 
    are juxtaposed for $N_c=2$, $\NfD=1$ and $G\Luv^2=6.8$\,. In both figures, the phase transition is second order.    
  }
\end{figure}

As $B_2(x)=x^2-x+1/6$ changes sign at $x=(3\pm \sqrt{3})/6$, it is useful to divide $0\leq \phi\leq \pi$ into two intervals.  
\begin{itemize}
  \item $\big(1-\frac{1}{\sqrt{3}}\big)\pi < \phi \leq \pi$:~ $B_2(\frac{\phi}{2\pi})<0$. For sufficiently small $L$, 
  chiral symmetry is restored. The chiral transition line on $L^{-1}$--$\phi$ plane is given by
  \begin{align}
    \frac{1}{\Luv L} & = \sqrt{ \left| B_2\left(\frac{\phi}{2\pi}\right) \right|^{-1} \!\! \left(\frac{1}{4\pi^2}-\frac{1}{2(N_c^2-1)G\Luv^2}\right) }\,,  
  \end{align}
  which up to typos confirms the result of Ref.~\cite[eq.~(6)]{Benic:2013zaa}. 
  \item $0\leq \phi \leq \big(1-\frac{1}{\sqrt{3}}\big)\pi$:~ $B_2(\frac{\phi}{2\pi})\geq 0$. 
  Chiral symmetry is spontaneously broken at all circumferences, $0<L<\infty$.  
  It is intriguing that the threshold $\phi=\big(1-\frac{1}{\sqrt{3}}\big)\pi\simeq 0.42\pi$ is a pure number, independent of any parameter of the model. 
\end{itemize}
In Figure \ref{fg:njlphases} (left) we show the $N_c=2$ phase structure with the parameter choice of $G\Luv^2=6.8$ that 
satisfies the bound \eqref{eq:chsb}. The region with larger $\frac{1}{\Luv L}$ is not shown 
because the model would be less reliable there owing to cutoff artifacts.%
\footnote{The absence of chiral restoration for $0\leq \phi \leq \big(1-\frac{1}{\sqrt{3}}\big)\pi$ is 
presumably a model artifact and not a genuine property of {\qcdad} because the gauge coupling  
runs to zero at $L\LQCD\ll 1$ and a spontaneous symmetry breaking should be hindered. 
This idea is consistent with the observation that a divergence of $T_c$ 
also occurs in a non-relativistic effective model with a cutoff \cite{Braun:2012ww} whereas it does not occur 
in the renormalizable Gross-Neveu model in 2 dimensions \cite{Karbstein:2006er}, with twisted boundary conditions, respectively.}  
The tendency that the chiral transition moves to higher $L^{-1}$ for smaller $\phi$ 
is clearly visible in the figure, which occurs \emph{within} the domain of validity of the model ($L^{-1}\lesssim \Luv$). 
A simplistic way of explaining this is to posit that a smaller $\phi$ which decreases the lowest Matsubara   
frequency ($\phi/L$) of fermions would facilitate symmetry breaking at low energy. 
We expect that the same behavior will be seen in actual {\qcdad}, in a phase with trivial holonomy.

\subsubsection{Center-symmetric holonomy}
\label{sec:confinedholo}

\begin{figure}[t]
  \centerline{
    \includegraphics[width=.85\textwidth]{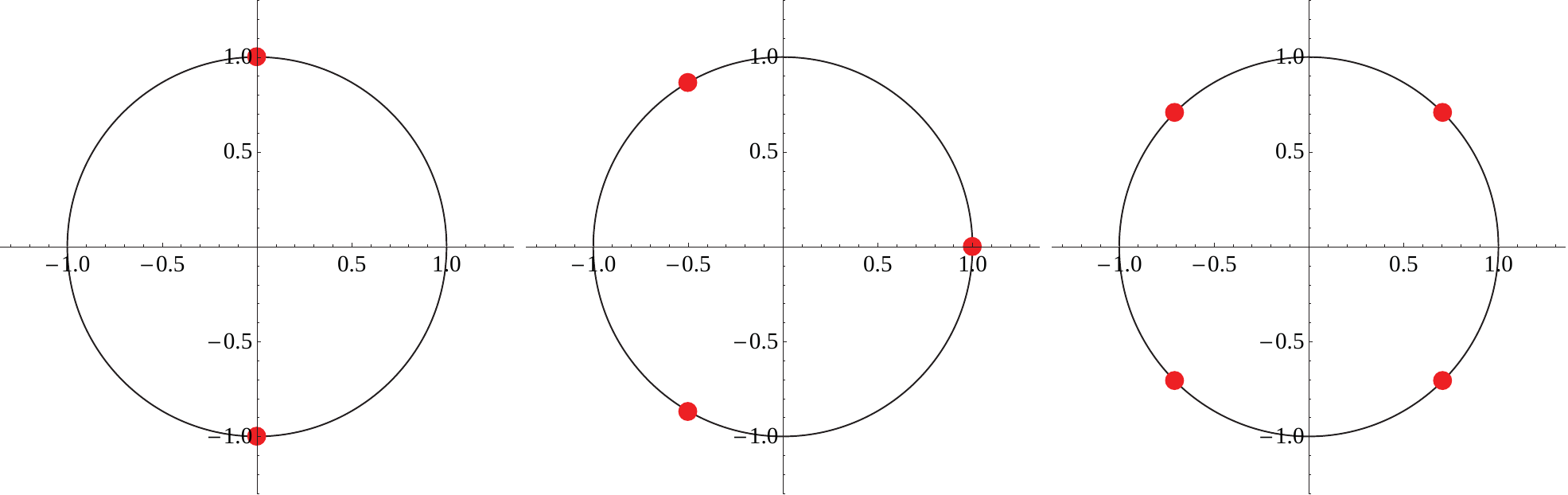}
    \put(-367,105){$N_c=2$}
    \put(-245,105){$N_c=3$}
    \put(-120,105){$N_c=4$}
  }
  \caption{\label{fg:holon}
    Eigenvalues of a center-symmetric holonomy $\Omega$ for $N_c=2$, $3$ and $4$  
    represented as red blobs on a unit circle in the complex plane. 
  }
\end{figure}

Next, we consider a center-symmetric background field 
\begin{align}
  A_4 & = \frac{1}{L} \diag\left( \Big(-1 + \frac{1}{N_c}\Big)\pi, \Big(-1 + \frac{3}{N_c}\Big)\pi, \dots, 
  \Big(1 - \frac{1}{N_c}\Big)\pi \right)\,. 
\end{align}
The eigenvalues of $\Omega=\ee^{iA_4}$ are equally spaced on a unit circle as depicted in Figure \ref{fg:holon}.  
It is easy to check that
\begin{align}
  \Tr_{\adj}(\Omega^n) & = |\!\Tr(\Omega^n)|^2-1 = \left\{
  \begin{array}{ll}
    N_c^2-1 ~~& \text{for}~~n\equiv 0~(\text{mod}~N_c)
    \\
    -1 & \text{otherwise} 
  \end{array} \right. \,.
\end{align}
Using this property in \eqref{preadjc}, we obtain
\begin{align}
  {\cal V}_{\adj}(N_{c}, \NfD=1, L, \md, \phi; \{q\}) 
  = \frac{\md^2}{\pi^{2} L^{2}} \sum_{n=1}^{\infty} 
  \frac{K_2(nL\md)}{n^2} \ee^{in\phi}\big( N_c^2 \delta_{n,0}^{(\text{mod }N_c)} -1 \big) + \text{c.c.} 
  \,,
  \label{eq:Vconfholo}
\end{align}
with $\delta_{n,0}^{(\text{mod }N_c)}=1$ for $n\equiv 0~(\text{mod}~N_c)$ and $=0$ otherwise. 
With the aid of the identity
\begin{align}
  \delta_{n,0}^{(\text{mod }N_c)} = \frac{1}{N_c}\sum_{k=0}^{N_c-1}\ee^{in\frac{2\pi k}{N_c}}\,,
\end{align}
\eqref{eq:Vconfholo} can be cast into the form
\begin{align}
{\cal V}_{\adj}(N_{c}, \NfD=1, L, \md, \phi; \{q\}) 
  & = N_c \sum_{k=0}^{N_c-1} \left\{
    \frac{\md^2}{\pi^{2} L^{2}} \sum_{n=1}^{\infty} 
    \frac{K_2(nL\md)}{n^2} \ee^{in\big(\phi+\frac{2\pi k}{N_c}\big)} + \text{c.c.} 
  \right\}
  \notag
  \\
  & \quad - \left\{
    \frac{\md^2}{\pi^{2} L^{2}} \sum_{n=1}^{\infty} 
    \frac{K_2(nL\md)}{n^2} \ee^{in\phi} + \text{c.c.}
  \right\} \,. 
\end{align}
The expansion in terms of $\md$ readily follows from \eqref{eq:Vadexp}.  
Combined with \eqref{eq:Vchiexp} it yields
\begin{align}
  {\cal V}_{\rm tot}(N_c,\NfD=1, \phi, L, \md ; \{q\})\Big|_{O\big(\md^2 \big)}
  = \left[ 
    \frac{1}{2G}-\frac{N_c^2-1}{4\pi^2}\Luv^2
    + \frac{1}{L^2} \xi\left(N_c; \frac{\phi}{2\pi}\right) 
  \right] \md^2 \,,
\end{align}
with
\begin{align}
  \xi(N_c;x) & \equiv B_2\left(\kakko{x}\right) 
  - N_c \sum_{k=0}^{N_c-1} B_2\left(\kakko{x+\frac{k}{N_c}}\right)
  \,.
\end{align}
The behavior of $\xi(N_c;x)$ depends on $N_c$. For $N_c=2$, it crosses zero at $x=k/3$ for $k\in\ZZ$. 
\begin{itemize}
  \item 
  $0 < \phi< 2\pi/3$:~ $\xi(N_c;x)>0$. For sufficiently small $L$, chiral symmetry is restored. The critical line 
  is given by
  \begin{align}
    \frac{1}{\Luv L} & = \sqrt{\xi\left(N_c; \frac{\phi}{2\pi}\right)^{-1}\left(\frac{N_c^2-1}{4\pi^2} - \frac{1}{2G\Luv^2}\right)} \,.
  \end{align}
  \item 
  $2\pi/3 \leq \phi \leq \pi$:~ $\xi(N_c;x)\leq 0$. Chiral symmetry is always broken for $0<L<\infty$.
\end{itemize}
In Figure \ref{fg:njlphases} (right) we show the $N_c=2$ phase structure with the parameter choice $G\Luv^2=6.8$ again. 
In comparison to the case with trivial holonomy, the locus of the chirally symmetric phase is now drastically different;  
it has been shifted to smaller $\phi$.  The overall structure of the plots in Figure \ref{fg:njlphases} are 
independent of the specific value of $G\Luv^2$ as long as the condition \eqref{eq:chsb} is satisfied.   
We conclude that the coupling to the Polyakov loop plays an essential role for the chiral symmetry realization 
with twisted boundary conditions.  

In Section \ref{sec:numerics} we will treat the holonomy as a dynamical variable and compare the analytical predictions with numerical results.

\subsection{PNJL model: numerical results}
\label{sec:numerics}

We limit ourselves to $N_c=2$ and $\NfD=1$ in the chiral limit $m=0$ 
to make numerical computations easier. 
The model has three dimensionful parameters: $M$, $G$ and $\Luv$. 
As we lack inputs from lattice simulations, there is no unique way to fix these parameters. 
In this study we will try two choices, so that by comparison 
one can see the parameter dependence of model predictions. 
As the first set of parameters, we use 
\begin{align}
  G\Luv^2=6.8 \qquad \text{and} \qquad  \Luv = 20 M.  
  \label{eq:parachoice}
\end{align}
We note that $G\Luv^2$ in \eqref{eq:parachoice} is the same as used for Figure \ref{fg:njlphases}. 

\begin{figure}[t]
  \centerline{
    \includegraphics[width=.5\textwidth]{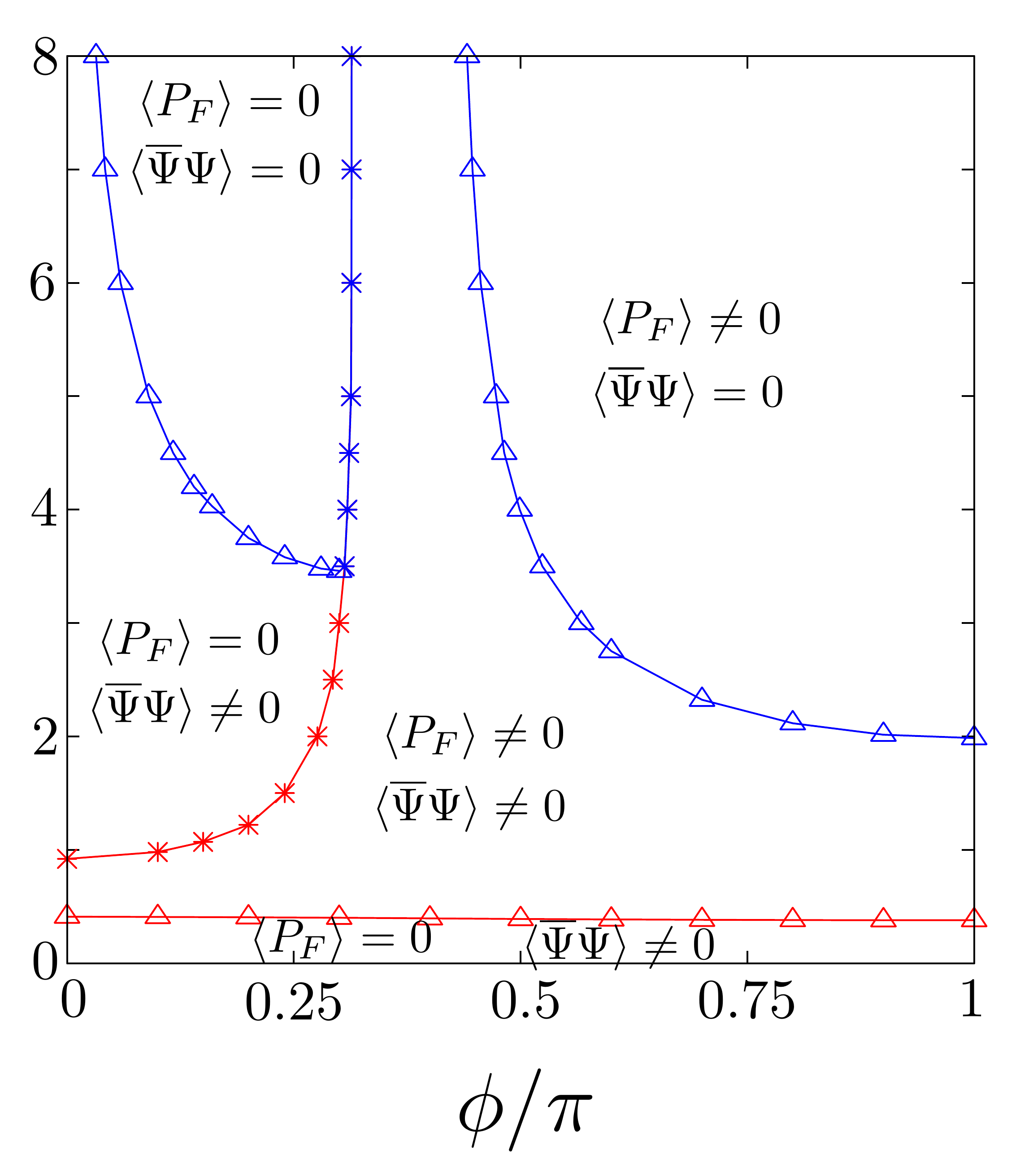}
    \put(-245,137){\Large$\displaystyle\frac{1}{LM}$}
  }
  \vspace{-.6\baselineskip}
  \caption{\label{fg:phdgm}
    A numerically obtained phase diagram of the PNJL model for $N_c=2$ and $\NfD=1$ in the mean-field approximation. 
    The blue (red) line denotes chiral (deconfinement) transition, respectively. The triangles ($\bigtriangleup$) represent a second-order transition 
    and the asterisks ($\ast$) a first-order transition.  
    The chiral and deconfinement transitions are degenerate for $\phi/\pi\simeq 0.3$ and $1/LM\gtrsim 3.4$. 
    We note that chiral condensate also jumps from nonzero to another nonzero value at the first-order deconfinement transition 
    (cf.~Figure \ref{fg:observ_L}).   
  }
\end{figure}

The phase diagram obtained from minimization of the thermodynamic potential \eqref{eq:Vtotal} 
for $0\leq 1/L \leq 8M$ and $0\leq \phi \leq \pi$ is displayed in Figure \ref{fg:phdgm}. It has some notable features: 
\begin{itemize}
  \item 
  The center phase structure at $1/LM\gg 1$ agrees with the one-loop analysis in \eqref{eq:table_2_1}, 
  including a center-changing transition at $\phi = 0.326\pi$. 
  \item 
  At $\phi=\pi$, the chiral and deconfinement transitions occur at widely separated temperatures: 
  we find $T_\chi\simeq 5 T_{\rm d}$, which is roughly consistent 
  with the interpolation of lattice data for $\NfD=1/2$ and $2$ in Ref.~\cite{Kogut:1986jt}. 
  This is comparable to {\qcdad} with $N_c=3$ and $\NfD=2$ 
  where $T_\chi\simeq 8 T_{\rm d}$ \cite{Karsch:1998qj,Engels:2005te}. 
  \item 
  The whole center phase structure of Figure \ref{fg:phdgm} is well captured by Figure \ref{phase1} with $m/M = 3.0$ 
  which is the \emph{bare} mass of fermions. This qualitative agreement is quite natural,  
  given that the \emph{dynamical} mass in the present setup, $\md\simeq 0.12 \Luv=2.4M$ at $1/LM=0$, is close to the value above.  
  The opening of a deconfined window for $0.5\lesssim \frac{1}{LM} \lesssim1$ at $\phi=0$ in Figure \ref{fg:phdgm} may be interpreted as follows: 
  the adjoint fermions with PBC, which favor a center-symmetric phase,  
  are suppressed by their large (bare or dynamical) mass and fail to prevent the appearance of 
  a center-broken phase at intermediate $L$.%
  \footnote{%
    At this point the reader may wonder why $\NfW=1$ adjoint fermion (called gluino) in ${\cal N}=1$ SYM can sustain   
    a center-symmetric phase at all $L$ despite that $\NfW=2$ ($>1$) adjoint fermions cannot. 
    It could be explained if the dynamical mass of gluinos in SYM were much smaller than the dynamical mass 
    originating from continuous chiral symmetry breaking for $\NfW=2$. This point deserves further study. 
  }   
  This phenomenon was observed for $N_c=3$ in lattice simulations
  at $\phi=0$ \cite{Cossu:2009sq,Cossu:2013ora} and confirmed in model calculation \cite{Nishimura:2009me}.  
  Moreover, Figure \ref{fg:phdgm} also reveals that the deconfined phase 
  extends to $0<\phi<\pi$ and separates the confining phase at small $1/L$ from 
  that at large $1/L$ altogether, in the entire phase diagram. This is a new result of this work.  
  \item 
  The lines of chiral phase transition in Figure \ref{fg:phdgm} 
  are well described by the analytic curves from the high-temperature expansion (Figure \ref{fg:njlphases}).  
  A notable difference, however, is that part of the critical line for center-symmetric holonomy (Figure \ref{fg:njlphases}, right) 
  is excised in Figure \ref{fg:phdgm} by a first-order deconfinement transition line. This is a manifestation of a nontrivial 
  interplay between chiral and center symmetry. 
  \item 
  Chiral symmetry is broken at all $0<L<\infty$ for $\phi=0$ and $0.326\pi \leq \phi \leq \big(1-\frac{1}{\sqrt{3}} \big)\pi$. 
  Actually, the fact that chiral symmetry is broken at $\phi=0$ up to a much smaller $L$ than at $\phi=\pi$ has been known  
  from lattice simulation \cite{Cossu:2009sq} and model calculations \cite{Nishimura:2009me,Kashiwa:2013rmg,Kouno:2013mma}, 
  whereas the behavior for $0.326\pi \leq \phi \leq \big(1-\frac{1}{\sqrt{3}} \big)\pi$ is a new finding here.  
  Although the model seems to capture the actual tendency of {\qcdad}, it has an intrinsic cutoff $\Luv$ and may not be trusted 
  at $1/L\gtrsim \Luv$; indeed, \"Unsal showed that chiral symmetry at $\phi=0$ is restored 
  at sufficiently small $L$ \cite{Unsal:2007vu,Unsal:2007jx}.  
\end{itemize}
\begin{figure}
  \centerline{
    \includegraphics[width=.44\textwidth]{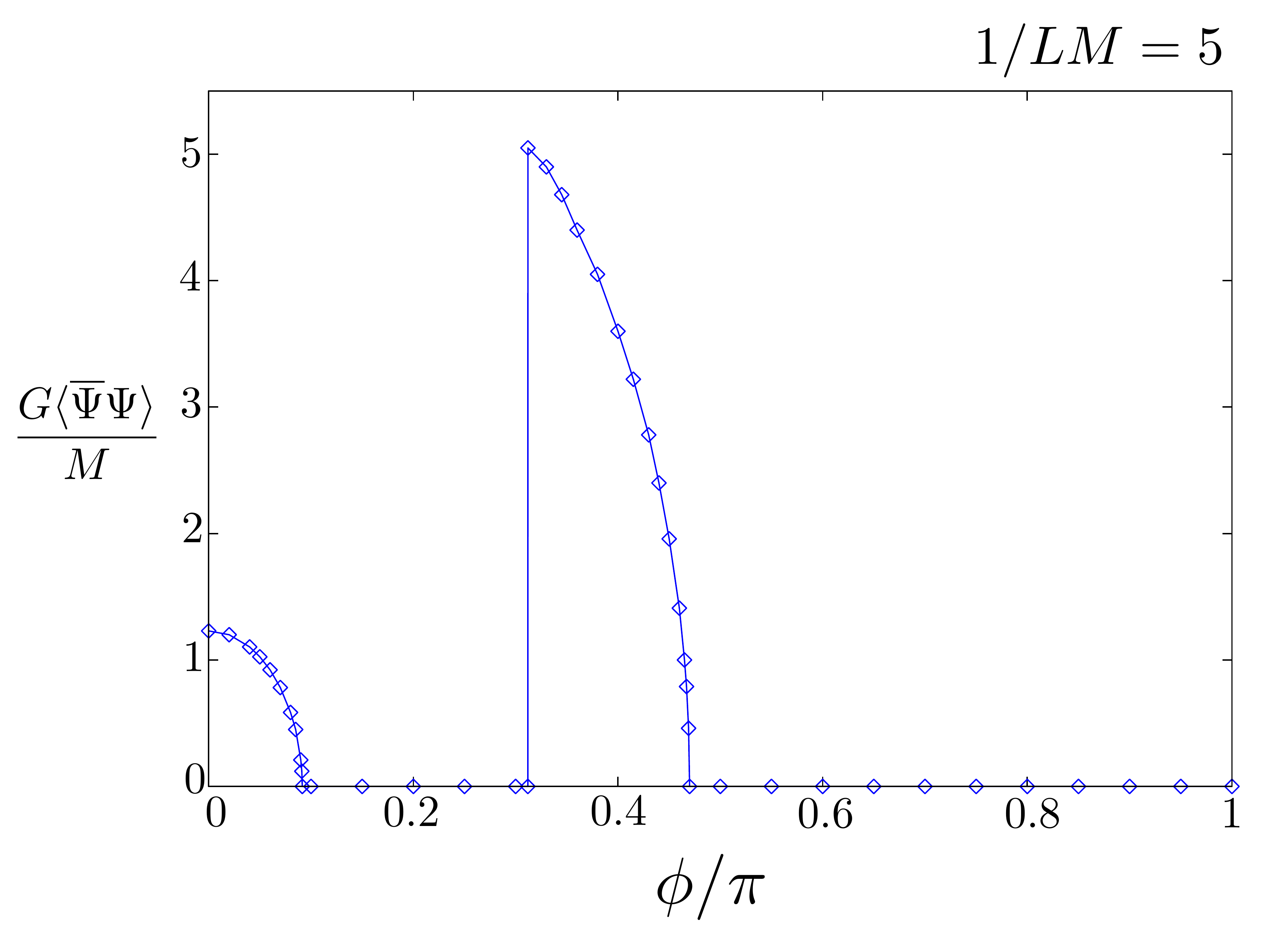}
    \quad 
    \includegraphics[width=.44\textwidth]{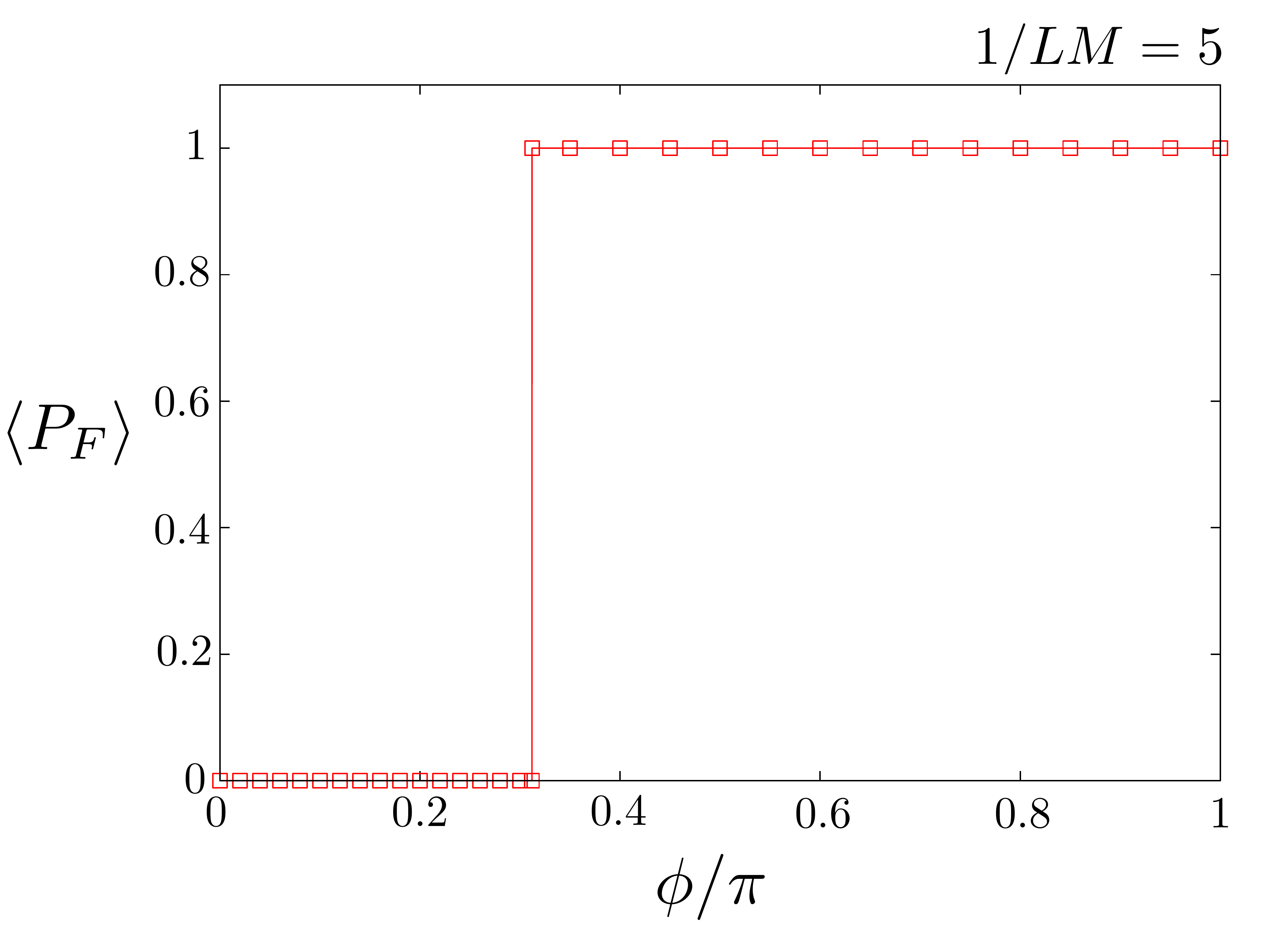}
  }
  \vspace{-.7\baselineskip}
  \caption{\label{fg:observ_phi}
    The $\phi$-dependence of the chiral condensate {\bf (left)} and 
    the Polyakov loop in the fundamental representation {\bf (right)} 
    at $1/LM=5$. 
  }
  \vspace{\baselineskip}
  \centerline{
    \includegraphics[width=.45\textwidth]{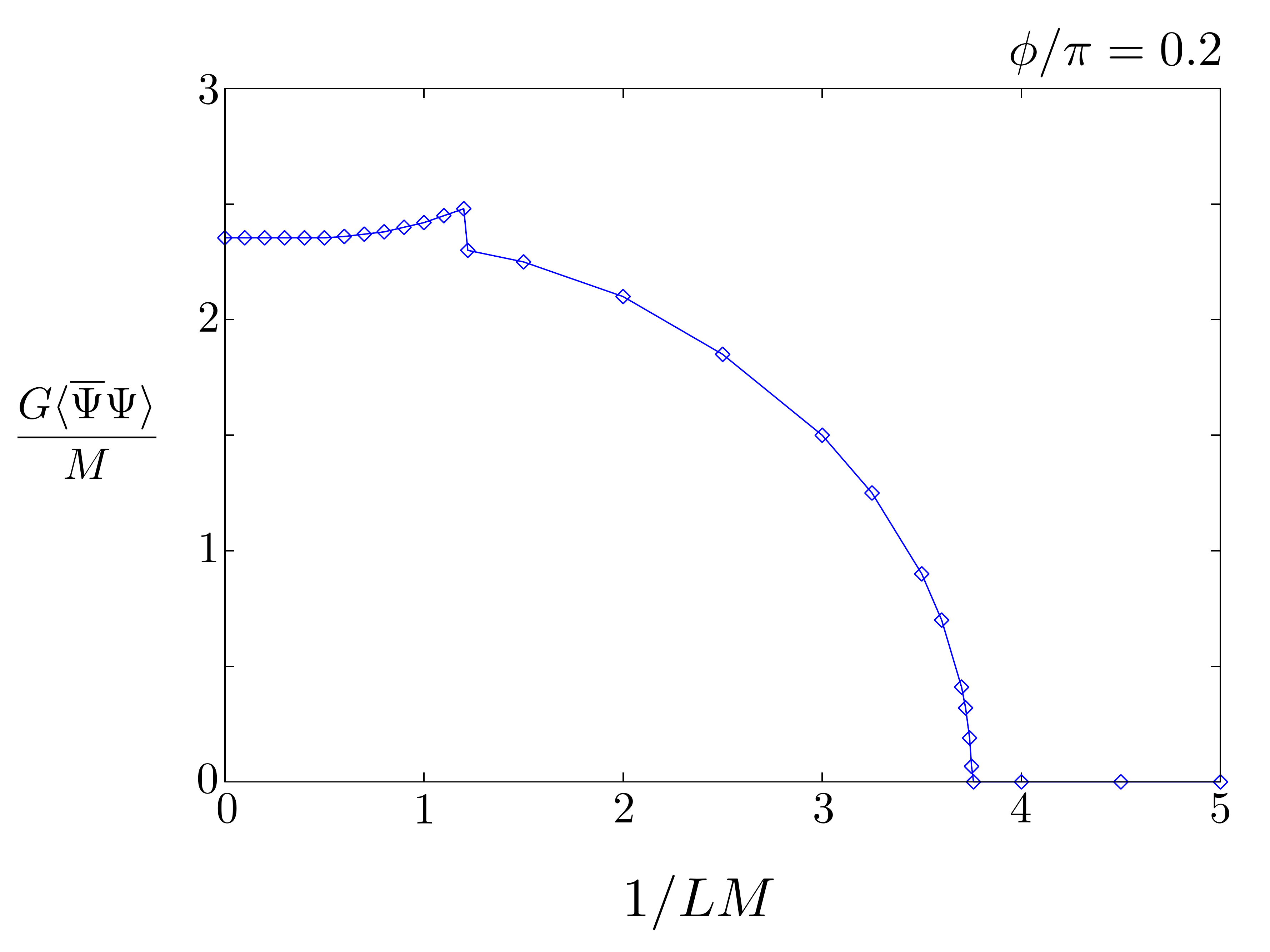}
    \quad \!\!\!\!
    \includegraphics[width=.45\textwidth]{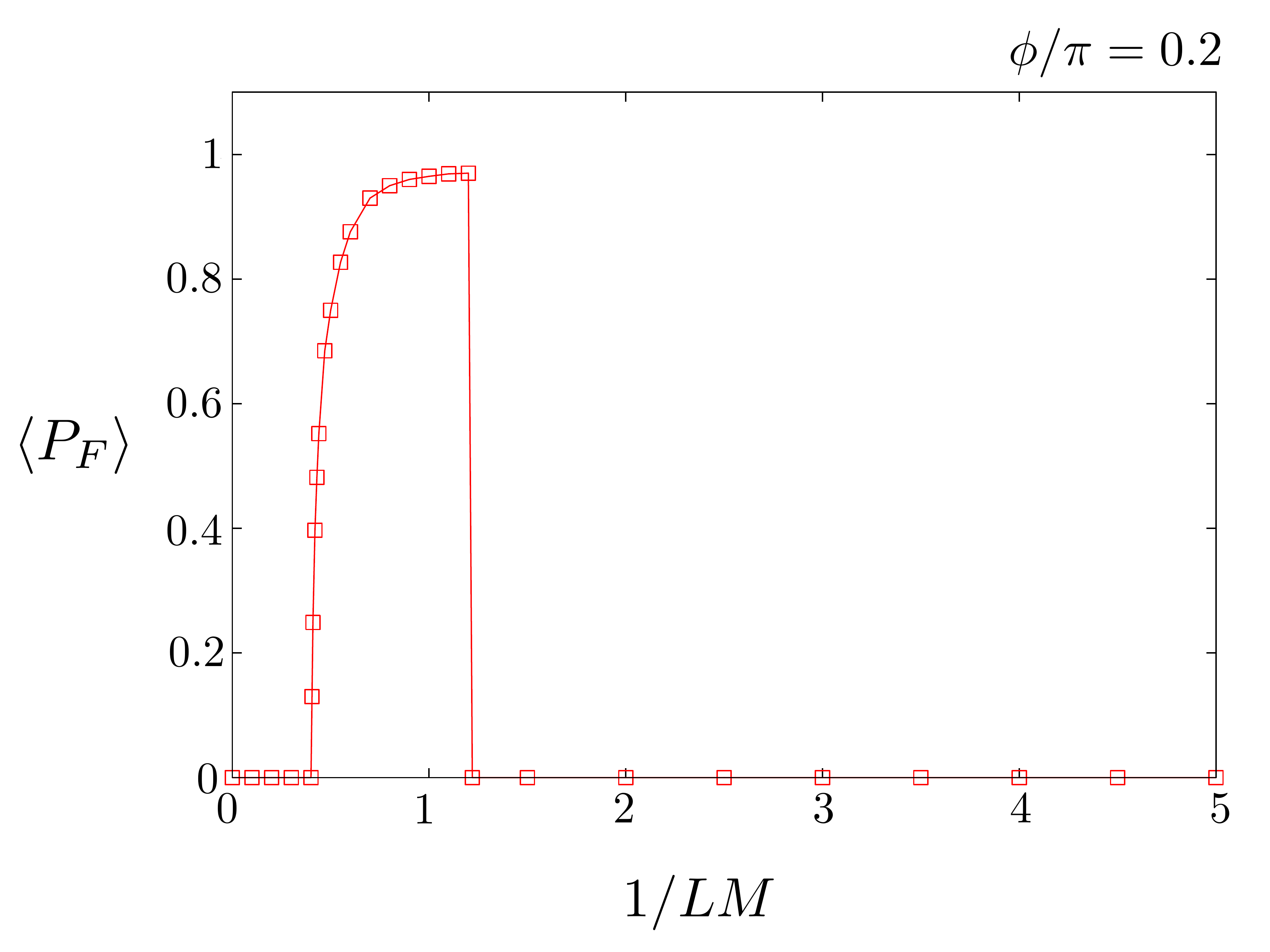}
  }
  \vspace{-.7\baselineskip}
  \caption{\label{fg:observ_L}
    The $L$-dependence of the chiral condensate {\bf (left)} and 
    the Polyakov loop in the fundamental representation {\bf (right)} 
    at $\phi/\pi=0.2$. 
  }
\end{figure}

In Figures \ref{fg:observ_phi} and \ref{fg:observ_L} we display 
the expectation values of the chiral condensate and the Polyakov loop 
as a function of $\phi$ and $L$. The result shows that the first-order 
transition in $\langle\PF\rangle$ is quite strong and is accompanied by 
a finite jump in the chiral condensate. 

\begin{figure}[t]
  \centerline{
    \includegraphics[width=.75\textwidth]{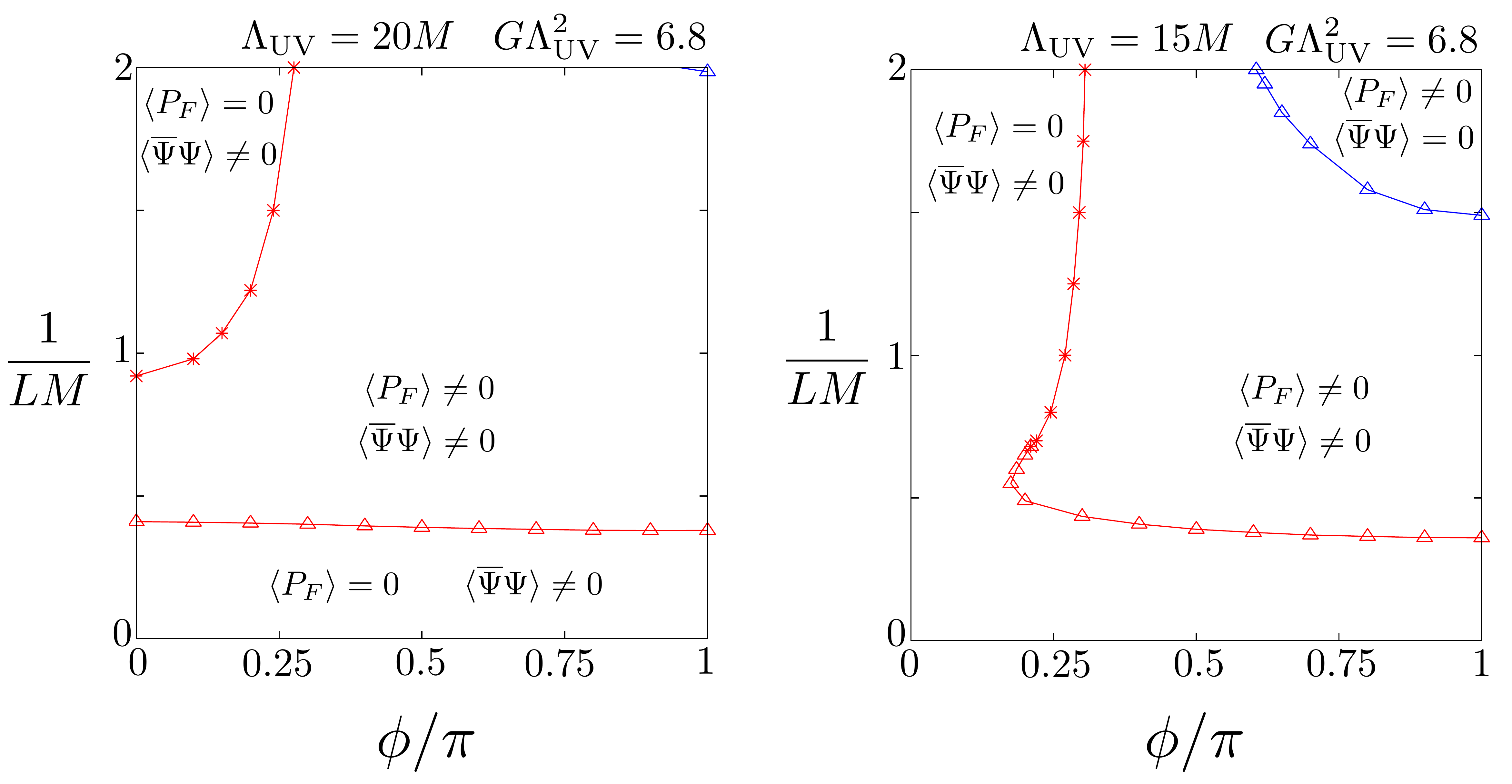}
  }
  \vspace{-.5\baselineskip}
  \caption{\label{fg:compareD}
    Phase diagrams of the PNJL model for $N_c=2$ and $\NfD=1$ with two different parameter sets,  
    \eqref{eq:parachoice} {\bf (left)} and \eqref{eq:parachoice2} {\bf (right)}, respectively. 
    The triangles ($\bigtriangleup$) represent a second-order transition 
    and the asterisks ($\ast$) a first-order transition.  
    The two confining phases are separated 
    in the left diagram while it is connected in the right diagram. 
  }
\end{figure}

As suggested by Figures \ref{phase1} and \ref{Phi_v}, it is the magnitude of $\md$ at low $1/LM$ that 
effectively determines the presence or absence of the deconfined phase on the $\phi=0$ axis.%
\footnote{$\md$ is roughly constant for $0 \leq 1/L \lesssim 2M$ as shown in Figure~\ref{fg:observ_L} (left).} 
In Figure \ref{fg:phdgm} the latter appears, because $\md =2.4M$ resulting from 
the parameter set \eqref{eq:parachoice} is greater than the threshold value, $1.878M$ (cf.~Figure~\ref{phase1}). 
Since the parameter fixing in the model is not unique, it is natural to ask whether 
the topology of the phase diagram could be altered or not with a different set of parameters. 
To address this issue, we also computed the phase diagram using the second set of parameters 
\begin{align}
  G\Luv^2=6.8 \qquad \text{and} \qquad  \Luv = 15 M \,,
  \label{eq:parachoice2}
\end{align}
leading to $\md=1.77M$, which is reduced from the previous case by 25\%. We note that 
$G\Luv^2$ in \eqref{eq:parachoice2} is not changed from \eqref{eq:parachoice}.     
For the transition temperatures at $\phi=\pi$, we found $T_\chi\simeq 4.5 T_{\rm d}$. 
The resulting phase diagram is presented in Figure~\ref{fg:compareD} (right), 
in comparison with that of the previous parameter set (left). 
We only show the region with $1/LM\leq 2$ because there is no qualitative difference between 
the two diagrams in the rest of the phase diagram.   

An important characteristic of Figure~\ref{fg:compareD} (right) is that 
the conventional confining phase on $\RR^4$ is now continuously connected 
to the other confining phase at large $1/LM$ where the gauge symmetry is spontaneously 
broken via the Hosotani mechanism. 
This is expected from Figures \ref{phase1} and \ref{Phi_v} since the dynamical 
mass here is lighter than the threshold value, i.e., $\md=1.77M < 1.878M$. 
The fact that the global topology of the phase diagram can be altered with such a small change 
of parameters ($\Luv = 15 M$ vs.~$\Luv = 20 M$) seems to highlight the crucial role of a 
delicate balance between dynamical mass $\md$ and the Yang-Mills scale $M$.   

\begin{figure}
  \centerline{
    \includegraphics[width=.5\textwidth]{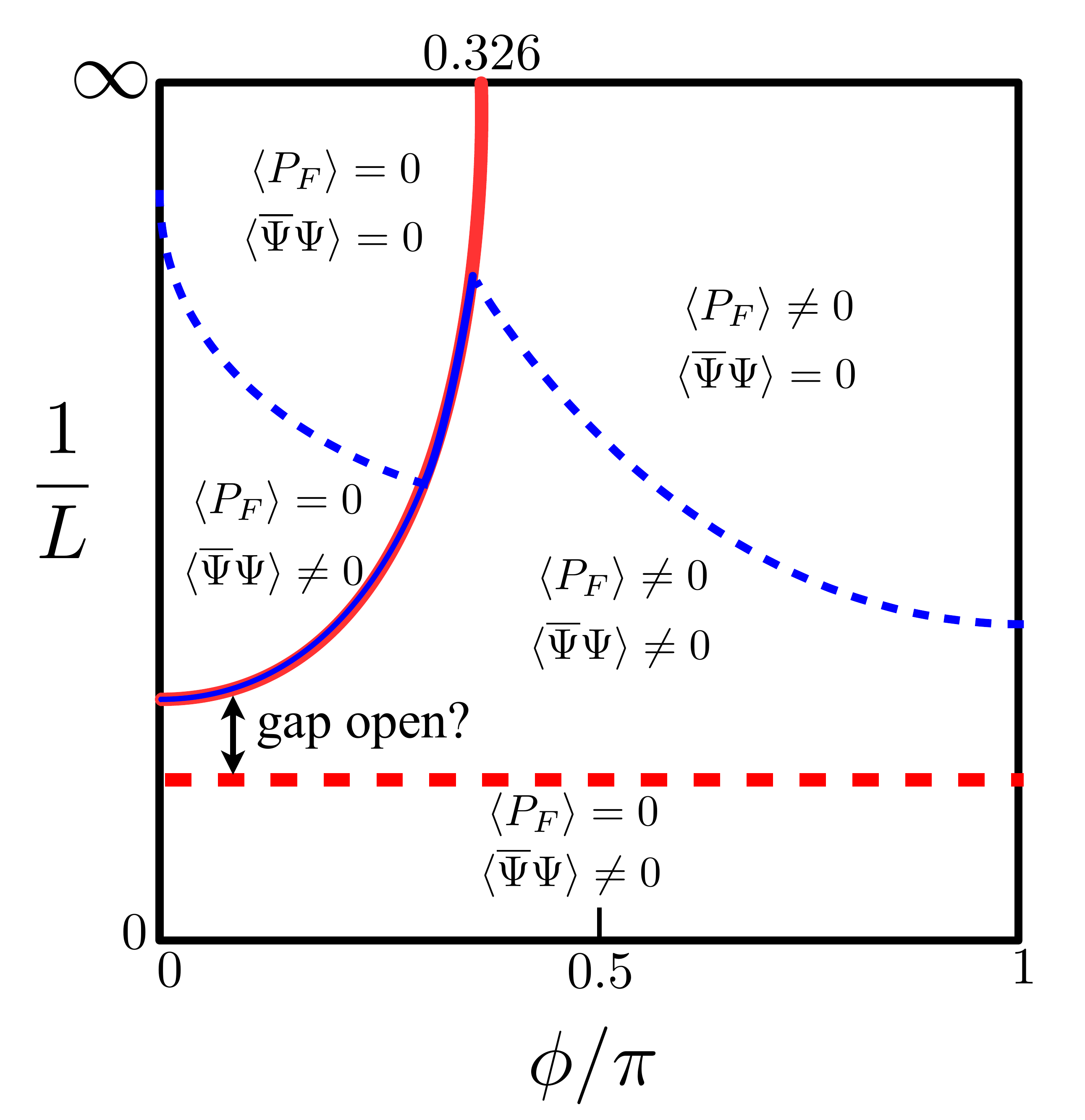}
  }
  \vspace{-.5\baselineskip}
  \caption{\label{fg:pdqcdadj}
    A conjectural phase diagram of {\qcdad} with $N_c=2$ and $\NfD=1$. 
    The red solid (dashed) line denotes a first-order (second-order) deconfinement transition. 
    The blue solid (dashed) line denotes a first-order (second-order) chiral transition. 
  }
\end{figure}

Finally, in Figure~\ref{fg:pdqcdadj} we show our conjecture for the phase diagram of {\qcdad} which is based on 
all insights from the model. The absence of chiral restoration for certain $\phi$ in the PNJL model is remedied by hand. 
The interesting prediction here is the merger of chiral and deconfinement transitions, indicated  
by a thin blue line overlaid on a red line in Figure~\ref{fg:pdqcdadj}. 
As for the center symmetry breaking at $\phi=0$, at present we do not know 
which phase diagram of Figure~\ref{fg:compareD} is the right one in {\qcdad}, so we have indicated 
our ignorance by the annotation ``gap open?'' in Figure~\ref{fg:pdqcdadj}. 
Comparing our conjectural phase diagram with an early conjecture by Shuryak \cite{Shuryak:2008eq}, we see several 
qualitative differences. Firstly, the lines of chiral and deconfinement transitions are assumed to intersect at a single point in Ref.~\cite{Shuryak:2008eq}, 
whereas our model calculation strongly suggests that they have an overlapping segment. Secondly, center symmetry at $\phi=0$ is 
assumed to be unbroken for all $0<L<\infty$ in Ref.~\cite{Shuryak:2008eq}, while this may not be necessarily true 
if the large constituent fermion mass is taken into account, 
as indicated in Figure \ref{fg:pdqcdadj}. 

The second-order chiral phase transition, represented by blue dashed lines 
in Figure \ref{fg:pdqcdadj}, characterizes the breaking/restoration of \emph{discrete} 
axial symmetry.  On the basis of the arguments in Section \ref{sec:prelim}, 
we expect that the chiral transitions on the right/left of the first-order 
deconfinement transition (red solid line in Figure \ref{fg:pdqcdadj}) 
will exhibit different universal behaviors: 
on the left side, it should belong to the 3d Ising universality class, while 
on the right side, to the 3d spin systems with $\ZZ_{4}$ symmetry breaking.

We hope that these predictions in this paper 
will be tested in future lattice QCD simulations.

\section{Summary and discussion}
\label{sec:SUM}

In this paper, we investigate QCD with adjoint fermions [{\qcdad}] on $\RR^3\times \SS^1$. 
In the past, {\qcdad} on $\RR^3\times \SS^1$ with the thermal boundary condition  
has been simulated on the lattice to test relationship between chiral symmetry breaking and confinement. 
{\qcdad} with the periodic boundary condition on $\SS^1$ has also been studied 
in the context of gauge symmetry breaking (the Hosotani mechanism \cite{Hosotani:1983xw}) and the semiclassical confinement  
due to bion condensation \`{a} la \"{U}nsal \cite{Unsal:2007jx}. However the other boundary conditions have not been 
systematically explored. In this work, we fill the gap by studying the non-perturbative 
dynamics of {\qcdad} on $\RR^3\times \SS^1$ with a generic boundary condition $0\leq\phi\leq\pi$ for fermions. 
We found a rich phase structure as a function of $\phi$, the fermion mass $m$, and the compactification size $L$. 

In Section \ref{sec:OL}, we examined the phase diagram at small $L$ by using one-loop 
perturbation theory for $\SU(2)$ and $\SU(3)$ gauge theories.  
We have shown that there is a critical boundary condition $\phi_c$ such that for $\phi_c<\phi<\pi$  
the center symmetry is spontaneously broken for all $0\leq mL\leq \infty$. 

We also studied the dynamics of BPS and KK monopoles associated with the gauge symmetry breaking 
$\SU(N)\to\U(1)^{N-1}$ by using semiclassical methods based on 
the index theorem (appendix \ref{app:indextheorem}) and found that, although 
they form molecules (\emph{bions}) at $\phi\ne 0$, their non-perturbative contribution to the mass gap and 
string tension is exponentially suppressed by a factor $\sim \ee^{-1/g}$ as compared to $\phi=0$, 
owing to the ``real mass'' of fermion zero modes that are exchanged between monopoles. 

In Section \ref{sec:NP}, employing a phenomenological model incorporating the Yang-Mills 
scale of confinement, we illustrated how the phase diagram for center symmetry 
evolves with varying $\phi$ and $m$ and interpolates the known three limits: $\phi=0$ (Hosotani-\"{U}nsal regime), 
$\phi=\pi$ (finite temperature) and $m=\infty$ (pure Yang-Mills). 

In Section \ref{sec:chisym}, we adopted the PNJL model to investigate the center and chiral phase structure 
as a function of $L$ and $\phi$, in the chiral limit $(m=0)$. 
We first performed a high-temperature expansion 
to grasp qualitative features of the phase diagram and showed that the background Polyakov loop 
strongly affects chiral symmetry realization at $\phi\ne 0$.  
Then we numerically solved the PNJL model. The result exhibits a rich phase structure, 
containing all four phases (with/without chiral/center symmetry breaking) which are separated by 
first- and second-order phase transition lines. 

One of the motivations of this work was to check the adiabatic continuity of center symmetry 
at $\phi=0$, posited in Refs.~\cite{Argyres:2012vv,Argyres:2012ka,Dunne:2012ae,Dunne:2012zk}. 
To address this issue, we adopted two parameter sets for the PNJL model and compared 
the resulting phase diagrams. For the first set, the $\U(1)^{N-1}$ confining phase at small $L$ 
is separated from the non-Abelian confining phase at large $L$ by a deconfined phase at intermediate $L$. 
For the second set, these two confining phases are continuously connected in the plane 
spanned by $L$ and $\phi$, especially on the $\phi=0$ axis. We find that which possibility is realized 
is determined by the magnitude of the constituent fermion mass: if it exceeds the confining scale of the 
Yang-Mills theory, they are separated, and if it is smaller, they get connected.%
\footnote{A similar conclusion was reached in Ref.~\cite{Nishimura:2009me} within the PNJL model at $\phi=0$.} 
Moreover, we also found that if the deconfined phase appears at intermediate $L$ on the $\phi=0$ axis, then 
it does not shrink but rather \emph{expands} at $\phi\ne 0$. This means that considering {\qcdad} 
in a two-dimensional $L$--$\phi$ plane does not help us rescue the adiabatic 
continuity, if it were not present at $\phi=0$.  

We hasten to add that, even if continuity did not exist between the two confining phases, 
it is not detrimental at all to the confinement mechanism via bion condensation and the resurgence framework for 
{\qcdad} at weak coupling \cite{Argyres:2012vv,Argyres:2012ka,Dunne:2012ae,Dunne:2012zk}. It however 
challenges the viewpoint that the elusive infrared renormalons on $\RR^4$ may be continuously 
related to semiclassical configurations (bions, bion--anti-bion molecules, etc.) on $\RR^3\times \SS^1$ 
with small $\SS^1$.%
\footnote{It should be noted that adiabatic continuity from small to large $L$ 
in {\qcdad} with $\NfW=1$ and $5$ stands on a solid ground; 
it is supersymmetry for $\NfW=1$ and infrared conformality for $\NfW=5$ \cite{Poppitz:2009uq} 
that ensures the absence of symmetry-changing phase transitions. 
However these are rather special theories and do not resemble QCD in the real world. 
The true problem posed here is whether continuity can hold in theories with continuous chiral symmetry breaking.} 
This is certainly an important problem and deserves further investigation. 

In the PNJL model, we have used a mean-field approximation to compute the expectation values 
of the Polyakov loop and the chiral condensate. 
This is expected to be reliable at least for weak coupling $L\LQCD\ll 1$ where the fluctuations of the order 
parameters are suppressed. It is also worth mentioning that a chiral model analysis with the mean-field 
approximation \cite{Nishimura:2009me} yields a phase diagram for {\qcdad} at $\phi=0$ 
which is in qualitative agreement with the lattice result for the entire $L$ range \cite{Cossu:2009sq}. 
Thus it seems reasonable to expect that the present work is also providing a qualitatively correct 
result for {\qcdad}, even though it is not theoretically warranted. As a future direction, we can 
employ a more refined scheme such as the Weiss mean-field approximation \cite{Zhang:2010kn} 
and the functional renormalization group \cite{Berges:2000ew} to check the phase diagram 
obtained in this work. 

It would be interesting to extend our non-perturbative analysis of the $N_c=2$ phase diagram 
to $N_c\geq 3$ and to the large-$N_c$ limit. The latter is of particular importance in the context of 
large-$N_c$ volume independence in {\qcdad} at $\phi=0$ \cite{Kovtun:2007py,Unsal:2010qh}. 
Detailed analysis of large-$N_c$ {\qcdad} at $\phi\ne 0$ may shed light on the nature of a novel fermionic 
symmetry in {\qcdad} recently claimed in Ref.~\cite{Basar:2013sza}. 

Our idea and methodology may be applied to the study of 
phenomenological extra-dimensional models beyond the standard model. 
Although the five-dimensional Hosotani mechanism with general twisted 
boundary conditions is known \cite{Hosotani:1988bm},
the phase structure in the space spanned by $\phi$, $m$ and $L$ 
has not been investigated in details.
Understanding of the phase diagram will help to find choices of parameters 
with desirable patterns of gauge symmetry breaking.
By combining analytic methods and lattice simulations to study the phase structure, 
we will be able to gain deeper understanding of the gauge-Higgs unification 
and other extra-dimensional models.


\acknowledgments 
The authors are grateful to E.~Itou and K.~Kashiwa for useful discussions and 
S.~Beni\'{c} for bringing Ref.~\cite{Benic:2013zaa} to their attention. 
Special thanks go to M.~\"{U}nsal for valuable comments on this work. 
The authors appreciate the KMI special lecture ``Resurgence and 
trans-series in quantum theories" at Nagoya University and thank the organizer T.~Kuroki. 
TM  is supported by the Japan Society for the Promotion of Science (JSPS)
Grants Number 26800147.  
TK was supported by the RIKEN iTHES Project and JSPS KAKENHI Grants Number 25887014.

\appendix

\section{Index theorem with twisted boundary conditions}
\label{app:indextheorem}

Here we study properties of the zero modes of the adjoint Dirac operator with a twisted boundary condition $\phi$. 
The main tool is the Nye-Singer index theorem for the Dirac operator on $\RR^3\times \SS^1$ \cite{Nye:2000eg,Poppitz:2008hr} 
which interpolates the APS index theorem on $\RR^4$ and the Callias index theorem on $\RR^3$. 
While it has been well known for the Dirac operator in the fundamental representation that the zero mode on a caloron 
with nontrivial holonomy ``jumps'' from one constituent monopole to another as $\phi$ is dialed \cite{Bruckmann:2003yq}, 
the behavior of adjoint zero modes has received less attention (but see Refs.~\cite{GarciaPerez:2006rt,GarciaPerez:2008gw,GarciaPerez:2009mg}) 
and it is the purpose of this appendix to summarize their properties as a background material for Section \ref{eq:semicla} in the main text.  

As argued in Ref.~\cite{Gross:1980br}, smooth finite-action gauge fields on $\RR^3\times \SS^1$ can be classified by 
the topological charge, the magnetic charge, and the holonomy at spatial infinity.  Let us assume that the holonomy at spatial infinity is given as
\begin{align}
  A_4\big|_{\infty} &= \frac{1}{L} \diag(q_1,q_2,\dots,q_N)  
  \quad \text{with}~~q_1<\dots<q_N \quad\text{and}~~\sum_{k=1}^{N}q_k=0\,,
\end{align}
where $L$ is the circumference of $\SS^1$. 
In this setup, monopoles of $N$ kinds are more fundamental topological objects than instantons, the latter being composed of the former. 
Now, we define the index $I^{(\phi)}_{\cal R}[n_1,\dots,n_N]$ of the Dirac operator in the representation ${\cal R}$, 
with a twisted boundary condition $\phi$ for fermions, as 
the number of right-handed normalizable zero modes 
minus the number of left-handed normalizable zero modes, in a background of $n_k$ monopoles of the $k$-th kind 
$(k=1,2,\dots,N)$. Here $n_N$ denotes the number of KK monopoles. 
Then, according to Ref.~\cite{Poppitz:2008hr}, the index for periodic boundary condition ($\phi=0$) is given by  
\begin{align}
  I^{(0)}_{\adj}[n_1,\dots,n_N] & = 2N n_N - \sum_{i,\, j = 1}^{N} \left\lfloor\frac{q_i-q_j}{2\pi}\right\rfloor \big\{(n_i-n_{i-1}) - (n_j-n_{j-1})\big\} \,,
\end{align}
with $\lfloor x \rfloor\equiv \max\{ k\in \ZZ \,|\, k \leq x \}$ and $n_0=n_N$ is understood. One can extend this formula to $\phi\ne 0$ by 
applying a constant  Abelian gauge field equal to $\phi/L$ along $\SS^1$. The result is 
\begin{align}
  I^{(\phi)}_{\adj}[n_1,\dots,n_N] & = 2N n_N - \sum_{i,\, j = 1}^{N} \left\lfloor\frac{q_i-q_j+\phi}{2\pi}\right\rfloor \big\{(n_i-n_{i-1}) - (n_j-n_{j-1})\big\} \,. 
  \label{eq:I_tw}
\end{align}
It is apparent from this expression that $I^{(\phi)}_{\adj}$ is periodic in $\phi$ modulo $2\pi$. It is also noteworthy that 
$I^{(\phi)}_{\adj}[1,1,\dots,1]=2N$, i.e.,  the index of BPST instanton is independent of $\phi$.  In what follows, we will examine the behavior of 
\eqref{eq:I_tw} for $N=2$ and $3$. 

\begin{figure}[t!]
  \centerline{
  \includegraphics[width=.48\textwidth]{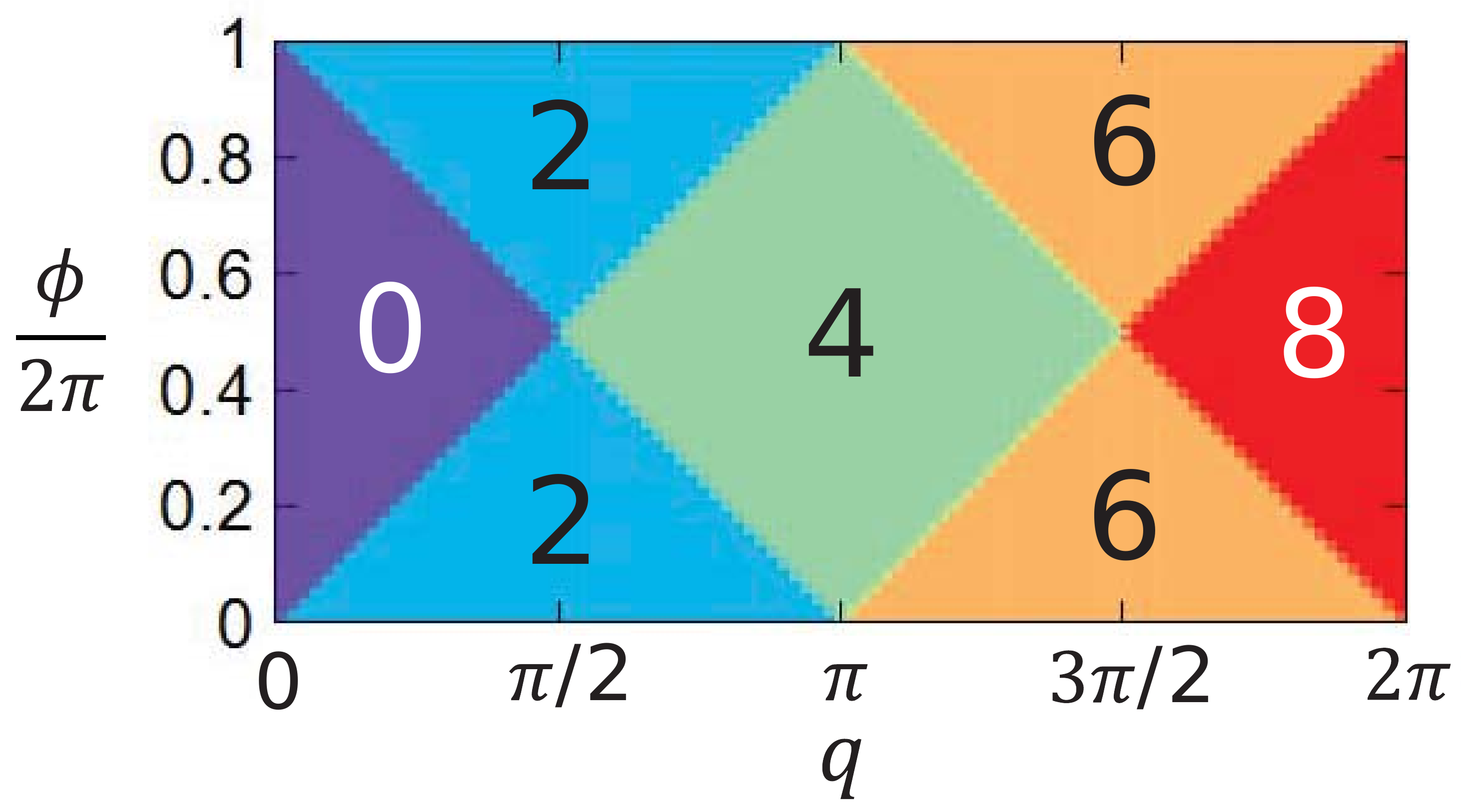}
  \quad 
  $\begin{matrix}
  \includegraphics[width=.36\textwidth]{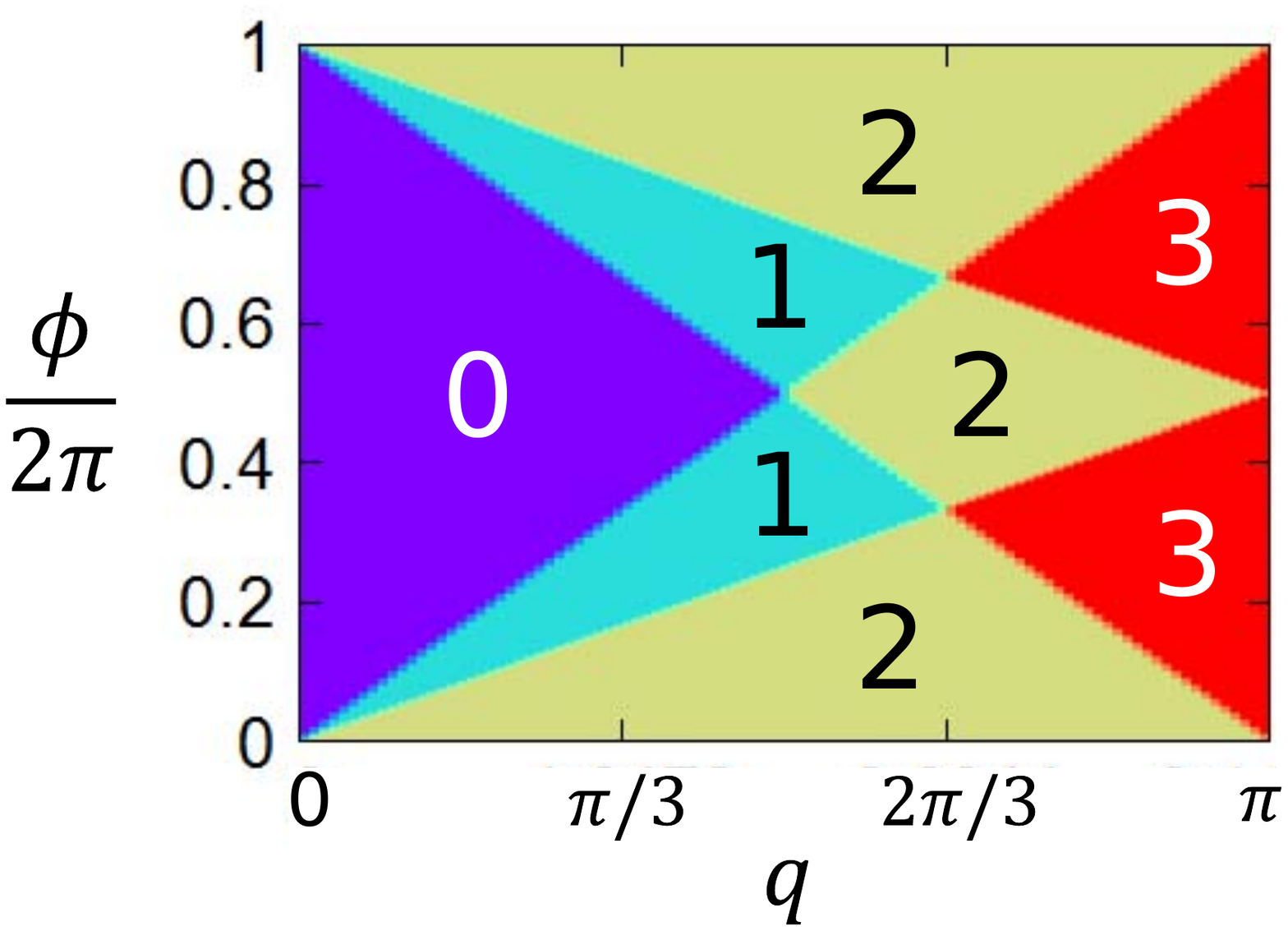}
  \vspace{91pt}\\
  \  
  \end{matrix}$
  }
  \vspace{-7.3\baselineskip}
  \caption{\label{fg:N2N3index} 
  Index of the adjoint Dirac operator in a monopole background:  
  $I^{(\phi)}_{\adj}[1,0]$ for $N=2$ {\bf (left)} and $I^{(\phi)}_{\adj}[1,0,0]$ for $N=3$ {\bf (right)}. 
  }
\end{figure}

For $N=2$, the holonomy may be parametrized as $A_4\big|_{\infty}=\frac{1}{L}\diag(-q,q)$ with $q>0$. Then 
\begin{align}
  I^{(\phi)}_{\adj}[1,0] & = 2 \left( \left\lfloor \frac{2q+\phi}{2\pi} \right\rfloor - \left\lfloor \frac{-2q+\phi}{2\pi} \right\rfloor \right)\,,
  \\
  I^{(\phi)}_{\adj}[0,1] & = 4 - I^{(\phi)}_{\adj}[1,0] \,. 
\end{align}
The contour plot of $I^{(\phi)}_{\adj}[1,0]$ is shown in Figure \ref{fg:N2N3index} (left). One can see that the index of a monopole indeed depends on 
the boundary condition. At $q=\pi/4$, for instance, the BPS monopole has two zero modes for $0\leq \frac{\phi}{2\pi} < 0.25$ as shown  
in the figure. When $\frac{\phi}{2\pi}$ exceeds $0.25$, they suddenly ``jump'' from the BPS monopole to the KK monopole. 
For $0.25 < \frac{\phi}{2\pi} < 0.75$ the KK monopole acquires all the four zero modes whereas the BPS monopole has none.  Finally, 
when $\frac{\phi}{2\pi}$ exceeds $0.75$, two of the zero modes ``come back'' to the BPS monopole.  

Also intriguing is the fact that the index does not change with $\phi$ when $q=\pi/2$, which corresponds to the center-symmetric background. 
This particular holonomy ensures that every monopole has two zero modes for all values of $\phi$. This finding is relevant to the 
semiclassical analysis on small $\SS^1$ in Section \ref{eq:semicla}.  

For $N=3$, for simplicity of exposition, we assume a specific holonomy 
$A_4\big|_{\infty}=\frac{1}{L}\diag(-q,0,q)$ with $q>0$. With a bit of algebra, one finds 
\begin{align}
  I^{(\phi)}_{\adj}[1,0,0] & = I^{(\phi)}_{\adj}[0,1,0] = 
  \left\lfloor \frac{q+\phi}{2\pi} \right\rfloor - \left\lfloor \frac{-q+\phi}{2\pi} \right\rfloor 
  + \left\lfloor \frac{2q+\phi}{2\pi} \right\rfloor - \left\lfloor \frac{- 2q+\phi}{2\pi} \right\rfloor 
  \,,
  \\
  I^{(\phi)}_{\adj}[0,0,1] & = 6 - 2 I^{(\phi)}_{\adj}[1,0,0] \,. 
\end{align}
The contour plot of $I^{(\phi)}_{\adj}[1,0,0]$ is shown in Figure \ref{fg:N2N3index} (right). Again one observes 
a nontrivial dependence of the index on $\phi$, but for $q=2\pi/3$ corresponding to the center-symmetric holonomy, 
the index is independent of $\phi$ and every monopole has two zero modes. 
We believe that this is true for general $N$: 
the index of every monopole should be equal to 2 for all $\phi$ in the center-symmetric holonomy
\begin{align}
  A_4\big|_{\infty} & = \frac{1}{L} \diag\left( \Big(-1 + \frac{1}{N}\Big)\pi, \Big(-1 + \frac{3}{N}\Big)\pi, \dots, 
  \Big(1 - \frac{1}{N}\Big)\pi \right)\,.
\end{align}
We have numerically verified this proposition for all $N\leq 10$ using \eqref{eq:I_tw}.

The behavior of anti-periodic zero modes ($\phi=\pi$) as a function of $q$ has already 
been analyzed in Refs.~\cite{GarciaPerez:2008gw,GarciaPerez:2009mg} 
on the basis of exact formulas for zero-mode densities. 
Their findings for $N=2$ and $3$ are totally consistent with Figure \ref{fg:N2N3index}.

\section{A remark on the literature}
\label{app:modeltable}

There are a number of preceding works based on chiral effective models for QCD and QCD-like theories. 
In this appendix we provide a summary of literature pertinent to the present work: specifically, 
in Table \ref{tb:models} we list up papers on chiral models  
that either involve adjoint fermions or impose twisted boundary conditions on fermions along $\SS^1$ 
(or, equivalently, involve imaginary chemical potential). 
Our work is also listed on the bottom row. The meaning of abbreviations is as follows. 

\begin{table}[t]
  \centerline{
  \scalebox{0.8}{
  \begin{tabular}{|c||c|c|c|c|c|c|c|}\hline
     & Boundary & Number & Chiral        & \multicolumn{2}{|c|}{\text{Rep.} of quarks} & \multicolumn{2}{|c|}{Behavior of ${\cal V}_\YM$} 
     \\ \cline{5-8}
     & condition  & of colors & interaction & fund. & adjoint              & large $\SS^1$ & small $\SS^1$ 
     \\ \hline \hline  
     $\begin{matrix}
       \textrm{Karbstein} \\ \textrm{\& Thies \cite{Karbstein:2006er} }
     \end{matrix}$ 
     & general & --- & GN$_2$ & --- & --- & --- & ---  
     \\ \hline 
     Sakai et al.~\cite{Sakai:2008py,Sakai:2008um} & general & 3 & NJL & \checkmark & --- & $\bigcirc$ & {\Large $\times$} 
     \\ \hline 
     Kashiwa et al.~\cite{Kashiwa:2009ki} & general & 3 & NJL & $\checkmark$ & --- & $\bigcirc$ & {\Large $\times$}
     \\ \hline 
     $\begin{matrix}{\rm Nishimura} \\ \textrm{\& Ogilvie~\cite{Nishimura:2009me}} \end{matrix}$ 
     & ABC/PBC & 3 & NJL & $\checkmark$ & $\checkmark$ & $\bigcirc$ & $\bigcirc$  
     \\ \hline
     Mukherjee et al.~\cite{Mukherjee:2010cp} & general & --- 
     & NJL & --- & --- & --- & ---  
     \\ \hline 
     Zhang et al.~\cite{Zhang:2010kn} & ABC & 2 and 3 & NJL$^{+}$ 
     & --- & $\checkmark$ & $\bigcirc$ & {\Large$\times$}
     \\ \hline 
     $\begin{matrix}{\rm Kashiwa} \\ \textrm{\& Misumi~\cite{Kashiwa:2013rmg}} \end{matrix}$ 
     & ABC/PBC & 3 & NJL & $\checkmark$ & $\checkmark$ & {\Large$\times$} & $\bigcirc$
     \\ \hline 
     Kouno et al.~\cite{Kouno:2013mma} & \scalebox{0.8}{$\begin{matrix} {\rm ABC/PBC/}\\{\rm FTBC}\end{matrix}$} & 3 & NJL & 
     $\checkmark$ & $\checkmark$ & $\begin{matrix}\text{$\bigcirc$ \scalebox{0.8}{(ABC/PBC)}}\\\text{{\Large$\times$} \scalebox{0.8}{(FTBC)}}\end{matrix}$ 
     & $\begin{matrix}\text{{\Large$\times$} \scalebox{0.8}{(ABC/PBC)}}\\\text{$\bigcirc$ \scalebox{0.8}{(FTBC)}}\end{matrix}$ 
     \\ \hline 
     Beni\'{c} \cite{Benic:2013zaa} & general & --- & NJL & --- & --- & --- & --- 
     \\ \hline 
     \textbf{\textsf{This work}} & general & 2 & NJL$^{+}$ & --- & $\checkmark$ & $\bigcirc$ & $\bigcirc$ 
     \\ \hline 
  \end{tabular}
  }}
  \caption{\label{tb:models}
    Chiral effective models with unorthodox fermionic boundary conditions (in chronological order). See the main text for the 
    definition of abbreviations. 
  }
\end{table}
\begin{center}
  \begin{minipage}{.95\textwidth}
    ABC: anti-periodic boundary condition, PBC: periodic boundary 
    condition, FTBC: flavor-dependent twisted boundary condition, GN$_2$: the Gross-Neveu model in 2 dimensions, 
    ${\cal V}_{\YM}$: the gluonic effective potential, NJL$^{+}$: the NJL model with the enlarged $\SU(2\NfD)$ symmetry of adjoint fermions.  
  \end{minipage}
\end{center}   
In the ``large $\SS^1$'' (``small $\SS^1$'') column, $\bigcirc$ is given if $V_{\YM}$ used in each work yields 
the confining phase at large $\SS^1$ (the perturbative one-loop potential for Polyakov-loop eigenvalues at small $\SS^1$), respectively. 
Otherwise $\times$ is given.

As is shown in the table, our work in this paper is the first study to consider generic twisted boundary conditions 
for adjoint fermions. We have used a four-fermi interaction derived in Ref.~\cite{Zhang:2010kn} to 
reflect the correct flavor symmetry of adjoint fermions. Another important ingredient of our model is 
the use of a gluonic potential ${\cal V}_{\YM}$ proposed in Ref.~\cite{Nishimura:2009me} which 
can produce a confining phase at large $\SS^1$ and a perturbative one-loop potential \eqref{gc} 
at small $\SS^1$. This property is essential for us to be able to describe 
the gauge-symmetry-broken phase at small $\SS^1$ and the confining phase at large $\SS^1$ 
in a unified manner.

\bibliography{Draft_pdflatex_v2.bbl}
\end{document}